\RequirePackage{fix-cm}
\documentclass[smallcondensed]{svjour3}     
\smartqed  
\usepackage{graphicx}
\usepackage{bbm,mathtools,amssymb}
\usepackage{hyperref}
\usepackage{pdflscape}

\usepackage{mathtools}
\newcommand{\assign}{\leftarrow}

\renewcommand{\d}{\Delta }
\newcommand{\x}{\mathbf{x}}

\newcommand{\prob}[1]{{#1}}
\newcommand{\R}{\ensuremath{\mathbb R}}

\newcommand{\D}{Z}

\DeclareMathOperator*{\argmin}{argmin}
\newcommand{\cS}{\mathcal{S}}
\newcommand{\SE}{\mathrm{SE}}
\newcommand{\MSE}{\mathrm{MSE}}
\newcommand\round[1]{\left[#1\right]}
\newcommand{\grad}{\nabla}

\begin{document}

\title{Four algorithms to solve symmetric multi-type non-negative matrix tri-factorization problem
}

\titlerunning{Four algorithms to solve symmetric multi-type NMTF problem}        

\author{Rok Hribar \and Timotej Hrga \and Gregor Papa \and Ga\v{s}per Petelin \and Janez Povh \and Nata\v{s}a Pr\v{z}ulj \and Vida Vuka\v{s}inovi\'{c}
}

\authorrunning{R. Hribar et al.} 

\institute{R. Hribar \at
              Computer Systems Department, Jo\v{z}ef Stefan Institute, Ljubljana, Slovenia \\
              Jo\v{z}ef Stefan International Postgraduate School, Ljubljana, Slovenia \\
              \email{rok.hribar@ijs.si}  *Corresponding author         
           \and
           T. Hrga \at
              University of Ljubljana, Faculty of Mechanical Engineering, Ljubljana, Slovenia\\
          \and
          G. Papa \at
              Computer Systems Department, Jo\v{z}ef Stefan Institute, Ljubljana, Slovenia \\
          \and
          G. Petelin \at
              Computer Systems Department, Jo\v{z}ef Stefan Institute, Ljubljana, Slovenia \\
              Jo\v{z}ef Stefan International Postgraduate School, Ljubljana, Slovenia \\
           \and
           J. Povh \at
              University of Ljubljana, Faculty of Mechanical Engineering, Ljubljana, Slovenia\\ 
              Institute of Mathematics, Physics and Mechanics, Ljubljana, Slovenia\\
           \and
           N. Pr\v{z}ulj \at
              Institute of Mathematics, Physics and Mechanics, Ljubljana, Slovenia\\
              Department of Computer Science, University College London, London, United Kingdom\\
              Barcelona Supercomputing Center, Barcelona, Spain\\
              ICREA, Barcelona, Spain\\
           \and
           V. Vuka\v{s}inovi\'{c} \at
              Computer Systems Department, Jo\v{z}ef Stefan Institute, Ljubljana, Slovenia \\
}

\date{Received: date / Accepted: date}

\maketitle

\begin{abstract}
In this paper, we consider the symmetric multi-type non-negative matrix tri-factorization problem (SNMTF), which attempts to    factorize several symmetric non-negative matrices simultaneously. This can be considered as a  generalization  of the classical non-negative matrix tri-factorization problem and includes a non-convex objective function which is a multivariate sixth degree polynomial and a has convex feasibility set. It has a special importance in data science, since it serves as a mathematical model for the fusion of different data sources in data clustering.

We develop four methods to solve the SNMTF. They are based on four theoretical approaches known from the literature: the fixed point method (FPM), the block-coordinate descent with projected gradient (BCD), the gradient method with exact line search (GM-ELS) and the adaptive moment estimation method (ADAM). For each of these methods we offer  a software implementation: for the former two methods we use Matlab and for the latter Python with the TensorFlow library.

We test these methods on three data-sets: the synthetic data-set we  generated, while the others represent real-life similarities between different objects.

Extensive numerical results show that with sufficient  computing time  all four methods perform satisfactorily and ADAM  most often yields the best mean square error ($\MSE$). However, if the computation time is limited,  FPM gives the best $\MSE$ because it shows the fastest convergence at the beginning.

All data-sets and codes are publicly available on our GitLab profile. 


\end{abstract}

\section{Introduction}

\subsection{Motivation}



With the expand of information technology the amount of data generated within different areas of our life has remarkably increased. As a consequence, the need to extract meaningful information out of the collected data has significantly grown and the knowledge discovery has become widely studied research area.
When we have different sources of data at disposal, a deeper insights can only come from a systems-level integration of all the available data. The major challenge is to design integrative and predictive heuristic computational methods that can collectively mine all the available data and can, as a result, enable the extraction of meaningful information.

Medicine and healthcare are exemplary areas where vast amounts of data has been collected and which attracts big attention of scientific community. Diverse molecular and clinical data describe various aspects of the human cells functioning and offer an unprecedented opportunity to revolutionize the biological and healthcare understanding.
Hence, methods which can enable the extraction of a discernible biological meaning from various multi-type, systems-level and high-throughput (so called ``omic'') networked molecular and clinical data \cite{PrDo:16} are needed.

Recently, a non-negative matrix factorization approach for network \emph{data integration} (also called \emph{data fusion}) was proposed \cite{GMDP:16a,GMDP:16,GMDP:16b,ZIT_et_al:13}. A new computational framework that is capable of harnessing this unprecedented opportunity provided by the wealth of data is based on a \emph{non-negative matrix tri-factorization} (NMTF) \cite{WTC:08,WHD:11}. In a nutshell, NMTF approximates a high-dimensional data matrix $R$, containing associations of $n_1$ data points of data type $1$ with $n_2$ data points of data type $2$, into a product of three, low-dimensional, non-negative matrix factors $R\approx G_1SG_2^\top$ by solving the following optimization problem:
\begin{align}
    \label{eqn:NMTF}  \tag{\prob{NMTF}}
    \min~  \{\|R-G_1SG_2^\top\|^2 \colon G_1,G_2,S\ge 0\},
\end{align}
where $\|\cdot\|$ is Frobenius norm in $\R^{n_1\times n_2}$.

The low-dimensional matrix factors $G_1$ and $G_2$ are used to assign $n_1$ data points into $k_1 \ll n_1$ clusters and $n_2$ data points into $k_2 \ll n_2$ clusters, respectively. This is known as the clustering property of the NMTF \cite{Detal:06}. The matrix $S$ is a $k_1\times k_2$ compressed representation of $R$ and  $G_1$ and $G_2$ imply \emph{co-clustering}. The reconstructed data matrix $G_1SG_2^\top$ is usually more complete then the initial data matrix $R$, featuring new entries, unobserved in the data, that emerged from the latent structure captured by the low-dimensional matrix factors. The observed property is known as the matrix completion property of the NMTF \cite{Cietal:09}.

\subsection{Problem formulation}

The central problem that we study in this paper is \emph{symmetric multi-type non-negative matrix tri-factorization (SNMTF)} problem. The input for this problem is an  $N$-tuple of symmetric non-negative matrices $(R_1,\ldots,R_N)$, which we want to factorize simultaneously by solving:
\begin{equation}
 \label{eqn:SNMTF} 
 \tag{\prob{SNMTF}}
\min\Big\{ \sum_{i=1}^N \|R_{i}-GS_{i}G^\top\|^2 : G,S_{i}\ge 0,~S_i=S_i^\top\Big\}.
\end{equation}

The \ref{eqn:SNMTF} can be considered as a specialisation of the NMTF since we have here $G_1=G_2$. On the other hand, we consider more than one input matrix simultaneously, so \ref{eqn:SNMTF} is in this sense also a generalization of \ref{eqn:NMTF}.
Problem \ref{eqn:SNMTF} is also a generalization of the symmetric non-negative matrix factorization problem \cite{huang2013non,lu2017nonconvex}, which can be formulated as:
   \begin{align}
  \label{eqn:SNMF}  \tag{\prob{SNMF}}
  \min~  \{\|R-GG^\top\|^2 \colon G\ge 0\}.
\end{align}
Generalisation goes into two directions: in \ref{eqn:SNMTF} we consider more than one matrix to be (simultaneously) decomposed and we allow a third factor $S$ to be sandwiched between the factors $G$ and $G^\top$. 
As a consequence
the existing methods for solving \ref{eqn:NMTF} and \ref{eqn:SNMF} are not straightforwardly applicable.

The \ref{eqn:SNMTF} is suitable for modelling the clustering problem with a single set of data points  and with different types of associations between them.
Each matrix $R_i$ contains one type of associations among the data points. This form is especially beneficial when there are different ways to measure associations, which complement each other. Important example is clustering of genes where several modalities of associations exist: gene co-expression, genetic similarity, protein-protein interaction, etc. Knowledge from each type of associations are combined to generate matrix $G$ that contains information about general properties of all associations, while condensed information about the different types of associations is captured in the matrices $S_i$.
For example,  the problem \ref{eqn:SNMTF} serves, after omitting the constraint $S_i\ge 0$, as the underlying mathematical model in the data-integrated cell in \cite{malod2019towards}.

In this paper we use two measures for factorization quality of feasible solutions for \ref{eqn:SNMTF}:
the \emph{square error} ($\SE$) and \emph{mean square error} ($\MSE$), as follows:
\begin{align}
    \SE = & \sum_i\|R_i-GS_iG^\top\|^2,\\
    \MSE = & \frac{\sum_i\|R_i-GS_iG^\top\|^2}{\sum_i\|R_i\|^2}.
\end{align}
The first measure is actually the objective function of \ref{eqn:SNMTF}, while the second is a relative value of the $\SE$ compared to the size of the input data.
We use  $\SE$ to define algorithms (gradients, steps sizes), while  $\MSE$ is used as one of the stopping criteria.

The existing complexity results show that \ref{eqn:SNMF} is NP-hard since the NP-hard problem whether given non-negative matrix is completely positive \cite{dickinson2014computational} can be reduced to solving \eqref{eqn:SNMF}.
The general non-negative matrix factorization problem is also NP-hard \cite{vavasis2009complexity}, but for the case of the \ref{eqn:SNMTF} the complexity is to the best of our knowledge not known. Nevertheless, the \ref{eqn:SNMTF} is a non-convex optimization problem with the objective function being a multivariate polynomial of degree 6, where the matrix
variables $G$ and $S_{i}$ must belong to the cones of non-negative matrices, so we conjecture that it is also in the class of NP-hard problems, which is supported by numerical evidence in this paper.

\vspace{-2mm}
\subsection{Our contribution}
\vspace{-1mm}

The central research question that we address in this paper is how \ref{eqn:SNMTF} can be efficiently solved with state-of-the-art mathematical optimization tools and methods. 


The main contributions of this paper are:
\begin{itemize}
\item We develop four algorithms to solve \ref{eqn:SNMTF}. They are problem-specific adaptations of four well-known methods: the \emph {fixed point method} (FPM), 
\emph{block-coordinate descent with projected gradient (BCD)}, \emph{gradient method with exact line search (GM-ELS)} and \emph{adaptive moment estimation method (ADAM)};
\item We develop and publish  efficient implementations of all four algorithms;
\item We provide and make publicly available three benchmark data-sets consisting of (i) synthetic matrices with known optimum solutions for \ref{eqn:SNMTF}, and (ii) real  data-sets from machine learning and (iii) from biology;
\item We provide extensive numerical evaluations of our code on these data-sets, which are the bases for our recommendations on which algorithm (and code) to  use to solve \ref{eqn:SNMTF} on given data-set.
\end{itemize}



The layout of the paper is as follows. A literature review on the topic is given in Section  \ref{sec:related_work}.
 Section \ref{sec:KKT} presents adaptations of FPM and BCD for \ref{eqn:SNMTF}, while GM-ELS and ADAM  are presented  in Section \ref{sec:ELS_ADAM}. 
The implementations of the proposed algorithms are described in Section \ref{sec:implementation} and analysed in Section \ref{sec:results}.
The implications of this work are discussed in 
Section \ref{sec:discussion}.

\section{Related work}
\label{sec:related_work}
Although \ref{eqn:SNMTF} can be treated as an exciting example of non-linear optimization, the mathematical optimization community did not invest much effort in solving it. Several researchers have studied other formulations of non-negative matrix tri-factorization problem and how to solve them using the fixed point method (FPM) with a different variant of the multiplicative updates.

The literature search shows that non-negative matrix factorization got its visibility and popularity mainly because of data science community, especially because of the clustering-like interpretations of non-negative factors.
Most of the results pertain to the classical (2-factor) non-negative matrix factorization problem, which can be formulated as:
  \begin{align}
  \label{eqn:NMF}  \tag{\prob{NMF}}
  \begin{split}
  \min~  \{\|R-GF^\top\|^2 \colon G,F\ge 0\}.
 \end{split}
\end{align}

The dissertation \cite{ho2008nonnegative} reviews different algorithms for solving \ref{eqn:NMF} and its generalizations. As an alternative, it suggests Rank-one residue iteration, an algorithm of low complexity which is reported to have fast convergence. The book \cite{cichocki2009nonnegative} contains comprehensive overview of results up to the year 2009. Another relevant survey with reviews of some standard algorithms for \ref{eqn:NMF} can be found in \cite{gillis2014and}. Other optimization methods applied to the non-negative matrix-factorization problems can be found in Lee and Seung \cite{LS:01}, where a diagonally rescaled gradient method for basic NMF problem is studied. Lin \cite{Lin:07} studied how to solve \ref{eqn:NMF} using a (projected) gradient method. An alternating, non-negative, least-squares algorithm with the active-set method is presented in \cite{KP:08}.

In \cite{GliJaPr:14,GMDP:16a,GMDP:16,GMDP:16b} big effort to network data fusion using non-negative matrices has been invested, but only relaxations of the \ref{eqn:NMTF}
and \ref{eqn:SNMTF} have been studied and only FPM was used.
In \cite{malod2019towards} the authors used a relaxation of the \ref{eqn:SNMTF} obtained by omitting the constraints $S_i\ge 0$ to develop
computational model called iCell which integrates different molecular interaction network types and provides new biological and medical insights related to cancer.

In \cite{Zitnik2019} the authors compared different approaches for solving the \ref{eqn:NMTF} problem on six large data-sets. Comparing alternating least squares, projected gradients, and coordinate descent methods they concluded that methods based on projected gradients and coordinate descent converge up to twenty-four times faster than multiplicative update rules and coordinate descent-based NMTF converges up to sixteen times faster compared to well-established methods.

In \cite{Buono2015,LiuWang2018} authors studied the \ref{eqn:NMTF} problem with orthogonality constraints. In \cite{Buono2015} the authors suggested a process of identifying a clearer correlation structure represented by the block matrix on two different NMF algorithms. The results showed that in most cases, the quality of the obtained clustering increases, especially in terms of average inter-cluster similarity. In \cite{LiuWang2018} the authors propose an L1-norm symmetric non-negative matrix tri-factorization method to solve the high-order co-clustering problem that aims to cluster multiple types of data simultaneously by utilizing the inter- or/and intra-type relationships across different data types. Due to orthogonal constraints and symmetric L1-norm formulation they derived the solution algorithm using the alternating direction method of multipliers.

Ding et al.\ \cite{Detal:06} studied several versions of tri-factorization problems including a problem similar to the \ref{eqn:SNMTF}, however, 
only the problem with one data matrix is studied and only FPM is used to obtain solutions that might not even be local optima. Wang et al. \cite{Wang_et_al:13} and \v{Z}itnik et al.~\cite{ZIT_et_al:13} studied a simplified version of the \ref{eqn:SNMTF}, with only one data matrix $R$. Again they used only the most straightforward formulation of FPM to obtain solutions close to a local minimum and to perform network-data fusion to predict the gene function. Another similar study can be found in \cite{WHD:11}, where a multi-type relational data is mined by solving a simplified version of the \ref{eqn:SNMTF}. These authors again used only FPM with adapted variants of the updating rules.

Non-negative matrix factorization approach can be found also in the evolutionary clustering. In \cite{YUWangJiaoLi2019} authors proposed a framework of evolutionary clustering based on graph regularised non-negative matrix factorization, to detect dynamic communities and the evolution patterns and to predict the varying structure across the temporal networks. The proposed approach has better performance on community detection in temporal networks compared to some widely used models based on evolutionary clustering and heuristic methods.
In \cite{Wang2011TowardsEN} evolutionary non-negative matrix factorization was proposed, which incrementally updates the factorized matrices in a computation and space efficient manner with the variation of the data matrix. 
In \cite{MaDong2017} they proposed semi-supervised evolutionary non-negative matrix factorization approach for detecting dynamic communities. The main advantage of the proposed approach is to escape the local optimum solution without increasing time complexity.
In \cite{Saito2015}, a time evolving non-negative matrix factorization was used for time-sequential matrix data to track the time-evolution of data clusters.

Bayesian semi-non-negative matrix tri-factorization method to identify associations between cancer phenotypes e.g., molecular sub-types or immunotherapy response, and pathways from the real-valued input matrix, e.g., gene expressions is proposed in \cite{Park2017,Park2020}.


\section{Fixed point method and block-coordinate descent approach to \ref{eqn:SNMTF}}
\label{sec:KKT}

In this section, we describe two out of the four methods which we devise to solve \ref{eqn:SNMTF}: the fixed point method (FPM) and the block coordinate descent with projected gradient (BCD). Both methods try to find a local minimum for the problem by addressing the Karush-Kuhn-Tucker (KKT) conditions. Note that the fixed point method could be also treated as a variant of block-coordinate descent method since in each iteration we make an update  only on one block of variables keeping the other fixed. However,  we decided to keep this notation  since it is most common in the literature.

\subsection{Karush-Kuhn-Tucker conditions of \ref{eqn:SNMTF}}
The KKT conditions consist of equations and inequalities that describe necessary conditions for local minima for the \ref{eqn:SNMTF}. The FPM and BCD methods try to find a solution for these conditions and this 
way a good approximation for a local minimum of \ref{eqn:SNMTF}. 

We start with the Lagrangian function for the \ref{eqn:SNMTF}:

\begin{align}
 \label{eqn:Lagrangian} 
 L(G,S,\beta,\gamma)~=~ \sum_{i}\|R_{i}-GS_{i}G^\top\|^2-
  \langle G,\beta \rangle- \sum_i\langle S_{i},\gamma_i\rangle.
\end{align}
Matrices $\beta$ and $\gamma_i$ are the Lagrangian multipliers (dual variables) for the non-negativity constraints $G\ge 0$ and $S_i\ge 0$, respectively.
By the definition of the \ref{eqn:SNMTF}, the matrices $R_i$ and $S_i$ are symmetric. Hence, Lagrangian multipliers $\gamma_i$ for $S_i$ are also symmetric and the Karush-Kuhn-Tucker conditions for the \ref{eqn:SNMTF} are:

\noindent\emph{Stationarity}
\begin{align}\footnotesize
\grad_G L ~=~ -4\sum_i R_i G S_i + 4 \sum_i G S_i G^\top G S_i - \beta ~ =~ & 0\label{eqn:kk1} \\
\grad_{S_i}L~=~ -2 G^\top R_i G
 +2 G^\top G S_i G^\top G - \gamma_i ~=~ & 0 \label{eqn:kk2}
\end{align}
\emph{Primal feasibility}
\begin{align}
G\ge 0,~S_i\ge 0,~\forall i \label{eqn:kk3}
\end{align}
\emph{Dual feasibility}
\begin{align}\label{eqn:kk4} 
\beta,~\gamma_i \ge 0,~\forall i 
\end{align}
\emph{Complementary slackness}
\begin{align}
\langle \beta, G \rangle~ =~ 0 &\label{eqn:kk5}\\
\langle \gamma_i, S_i \rangle~ =~ 0 &,~\forall i \label{eqn:kk6}
\end{align}

By $\grad_G L $ and $\grad_{S_i} L $ we denoted the gradients of $L$ in variables $G$ and $S_i$, respectively.

Since the Slater  constraint qualification is trivially satisfied for the constraints of  \ref{eqn:SNMTF}, it holds from the Lagrangian theory (see, e.g., \cite{Bertsekas-2016}) that any candidate for a local minimum of the \ref{eqn:SNMTF} must satisfy the necessary conditions which consist of \eqref{eqn:kk3}--\eqref{eqn:kk6} and:

\begin{align}
\beta = &\sum_i \Big(-4 R_i G S_i + 4 G S_i G^\top G S_i \Big) \label{eqn:kk1a} \\
\gamma_i =& -2 G^\top R_i G + 
2 G^\top G S_i G^\top G,~~\forall i \label{eqn:kk2a}
\end{align}

The FPM and BCD methods devised for the \ref{eqn:SNMTF} try to find a solution satisfying \eqref{eqn:kk3}--\eqref{eqn:kk2a} conditions.
Both have already been used for standard non-negative matrix factorization problems \cite{AsadiPovh:20,Detal:06,Mirzal2014,MirzalUnpublished} and in this paper we adapt them to work for the \ref{eqn:SNMTF}.

\subsection{Fixed point method (FPM)}

The fixed point method is a standard and the most common approximation method to solve different variants of the NMF problem.
It is reduced to the so-called multiplicative update rules, see, e.g., \cite{GliJaPr:14,WTC:08}. In the specific case of \ref{eqn:SNMTF}, conditions \eqref{eqn:kk3}--\eqref{eqn:kk2a} imply:

\begin{align}
G \odot \beta=&G \odot \sum_i \Big(-4 R_i G S_i + 4 G S_i G^\top G S_i \Big) = 0,\label{eq:compl1} \\
S_i \odot \gamma_i = & S_i \odot (-2 G^\top R_i G + 
2 G^\top G S_i G^\top G )= 0,~\forall i.\label{eq:compl2}
\end{align}

In the terms above we use $\odot$ to denote the Hadamard (element-wise) matrix product.
Since $G\ge 0,~S_i\ge 0,\forall i$, we can substitute $G \rightarrow G \odot G,~S_i\rightarrow S_i \odot S_i$ 
in \eqref{eq:compl1}--\eqref{eq:compl2}. Therefore, \eqref{eq:compl1}  is equivalent to 
\begin{equation*}\small 
     G\odot G  \odot \sum_i \Big(G S_i G^\top G S_i)^+-(G S_i G^\top G S_i)^-\Big)=G\odot G  \odot  \sum_i \Big((R_i G S_i)^+-(R_i G S_i)^- \Big),
\end{equation*}
where  $a=a^+-a^-$,  $a^+=\max(a,0) $,  $a^-=\max(-a,0)$, is the  standard decomposition of real numbers.

By rearranging the terms such that we have only positive terms on each side we obtain  
\begin{equation*}\small 
  G\odot G \odot \sum_i  \Big( (G S_i G^\top G S_i)^+ +(R_i G S_i)^- \Big) = G\odot G\odot   \sum_i \Big(  (G S_i G^\top G S_i)^- +(R_i G S_i)^+ \Big).
  \end{equation*}

After division with the term on the left and by applying the square root on both sides we obtain  the following multiplicative update rule for $G$:

\begin{align}
G & \leftarrow G\odot \sqrt{\frac{\sum_i\Big(  (GS_iG^\top GS_i)^-+(R_iGS_i)^+ \Big)}{\sum_i \Big((GS_iG^\top GS_i)^++(R_iGS_i)^-\Big)}},\label{eqn:MU1}
\end{align}
where the division and square root are applied element-wise. 
Similarly, we develop the update rule for $S_i$:

\begin{align}
S_i & \leftarrow S_i\odot
\sqrt{\frac{(G^\top G S_i G^\top G )^-+(G^\top R_i G)^+ }{(G^\top G S_i G^\top G )^+ +(G^\top R_i G)^-}}.\label{eqn:MU2}
\end{align}

 We could use also the reverse order - division with the term on the right, but this yields in practice worse convergence.

Note that if we start with non-negative $G$ and $S_i$, 
then we retain their non-negativity throughout the iterations, and consequently the products of 
matrices in \eqref{eqn:MU1}--\eqref{eqn:MU2} are non-negative (recall, $R_i$ are assumed to be non-negative), so all terms $(\cdot )^-$ are zero and these multiplicative update rules  simplify to 

 \begin{align}
 G \leftarrow & ~G \odot \sqrt{(\sum_i R_iGS_i )\oslash (\sum_i GS_iG^\top GS_i) },\label{eqn:MU1a}\\
 S_i \leftarrow 	&~S_i\odot 
 \sqrt{(G^\top R_i G)
 		 \oslash
 		(G^\top G S_i G^\top G) },\label{eqn:MU2a}
 \end{align}
 where $\oslash$ denotes the element-wise division.
 The fixed-point method for the \ref{eqn:SNMTF} is summarised in Figure \ref{fig:FPM}.
 
 \begin{figure}
 \caption{Fixed point method to solve \ref{eqn:SNMTF}}
 \label{fig:FPM}
\fbox{\parbox{\textwidth}{
\begin{enumerate}
		\item[]{\bf Input:} Non-negative symmetric $R_1,\ldots,R_N$.
			\item[1.] 		{\bf Initialisation:}
			Compute initial non-negative matrices $G$ and $S_i$.			
		   \item[2.]{\bf While} termination test not satisfied 
		   \begin{itemize}
			 \item [2.1]  Compute  $G_{new}$ from current $G$ and $S_i$ using \eqref{eqn:MU1a}.
             \item [2.2]  {\bf For} $i = 1, 2,\ldots, N$
             \begin{itemize}
                \item [2.2.1] compute  $(S_i)_{new}$  from  $G_{new}$ and current $S_i$ using \eqref{eqn:MU2a}.
            \end{itemize}
		   \end{itemize}
	\end{enumerate}
}}
\end{figure}

\subsection{Block-coordinate descent method with projected gradient method (BCD)}\label{sec:BCD_PGM}

\begin{figure}
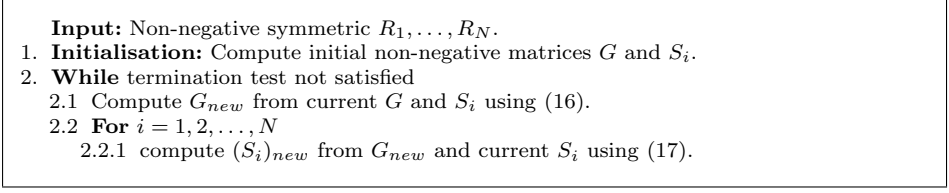
\caption{2-block coordinate descent algorithm}\label{fig:1}
\fbox{\parbox{\textwidth}{
\begin{enumerate}
		\item[]{\bf Input:} Non-negative symmetric $R_1,\ldots,R_N$.
			\item[1.] 		{\bf Initialisation:}
			Compute initial values for $G$.			
		   \item[2.]{\bf While} termination test not satisfied 
		   \begin{itemize}
			 \item [2.1]	Compute $$(S_1,\ldots,S_N)_{new}=\argmin\{\sum_{i=1}^N {\Vert  R_i -GS_iG^\top\Vert}^2  \mid S_i \ge 0, ~S_i=S_i^\top\}.$$
             \item [2.2] $(S_1,\ldots,S_N)=(S_1,\ldots,S_N)_{new}$.
             \item [2.3]  $$G_{new}=\argmin\{\sum_{i=1}^N {\Vert  R_i -GS_iG^\top\Vert}^2  \mid G \ge 0\}.$$
             \item [2.4] $G=G_{new}$.
		   \end{itemize}
	\end{enumerate}
}}
\end{figure}

In this subsection we show how to apply to the \ref{eqn:SNMTF} a well-known block-coordinate descent method, also known as non-linear Gauss-Seidel method, see e.g. \cite{Bertsekas-2016,tseng2001convergence,wright2015coordinate}. We can consider \ref{eqn:SNMTF} as an optimization problem in $(1+N)$-block variables $(G, S_1,S_2,\ldots,S_N)$ that must belong to the cone $\R^{n\times k}_+\times (\cS_k)_+^N$. 
We separate the feasible set into direct product of two cones, one consisting of $R^{n\times k}_+$ and the other consisting of $(\cS_k)_+^N$. We optimize $G$, while $\{S_i\}$ 
are fixed and vice versa. Note, that for fixed $G$, the \ref{eqn:SNMTF} problem decomposes into $N$ convex quadratic subproblems in nonnegative variables $S_i$. 
Therefore, we propose a 2-block coordinate descent algorithm, as described in Figure \ref{fig:1}.

The optimization problem in Step 2.1 in Figure \ref{fig:1} is ``easy'', which means that it boils down to solving $N$ convex quadratic optimization problems over the non-negative orthants. We have a large variety of iterative methods to solve them, including interior-point methods, projected gradient methods, active set methods etc., for a brief and a comprehensive overview on different special types of quadratic optimization problems, methods to solve them and state-of-the-art solvers see \cite{furini2019qplib}. We decided to use a simple projected gradient method with exact line search, since it is very easy to implement and it works very well. We tried also to use existing solvers for such problems, like the {\tt quadprog} function in Matlab \cite{MATLAB:2019} or {\tt mskqpopt} from MOSEK, but the time performance was much worse.
 
The problem in Step 2.3 is more involved. It is a non-convex programming problem since the objective function is multivariate polynomial of order 4 in matrix variable $G$.
We decided to use projected gradient method, where the step-length could be determined either using Armijo rule or by exact line search. After extensive testing we decide for the latter. 
For example, the exact line search for the update of $G$ in the direction of gradient $\d G= \grad_G \SE$ boils down to finding optimum $t$ such that the univariate polynomial of degree 4
$$p(t)=\sum_i\|R_i-(G+t\, \d G)S_i(G+t\, \d G)^\top\|^2$$ is minimum. 

We compute coefficients of $p(t)$ by writing them in closed form formulae and computing them exactly, using all available information, like symmetries of $R_i$ and $S_i$.
Note that we  also tried to compute the coefficients of $p(t)$   approximately by evaluating $p(t)$ on at least 5 points and using curve fitting methods, like {\tt polyfit} in Matlab. However, this method finds best curve in a least-squares sense, which in practice gives almost exact values for the coefficients of the highest degree terms (for $t^4$ and $t^3$) which were also the largest, while the  linear term coefficient   usually deviates a lot from the exact value. Since this coefficient plays an important role in computing the minimum of $p(t)$, which is usually close to 0, we decided not to use this approach. Similar behaviour was observed with the Gradient method with exact line search (GM-ELS)   using Python and TensorFlow  in Section \ref{sec:ELS_ADAM}.\label{note:polyfit}
 
Once we know $p(t)$, we compute its  minimum numerically by the Matlab function {\tt fminbnd} over the interval $[-1,0]$.
If $t_{\min} = 0$ or if $p(t_{\min})-p(0)>-10^{-3}$, we take 
$G_{new}=G+t_{\min}\d G+10^{-5}\cdot {\tt rand}(n,k)$, otherwise we take $G_{new}=G+t_{\min}\d G$, where 
${\tt rand}(n,k)$ is a random matrix with entries uniformly distributed on $(0,1)$.
Finally, we project  $G_{new}$ to the cone of non-negative matrices 
$G_{new}=\max(G_{new},0)$.

To compute new iterate $(S_1,\ldots,S_N)_{new}$, we again perform exact line search in the directions defined by gradients 
$\d S_i = \grad_{S_i}\SE$, but in this case we have for each $i=1,\ldots,N$
$$p_i(t)=\|R_i-G(S_i+t\,\d S_i)G^\top\|^2,$$
which is a convex quadratic polynomial of degree 2 in $t$ with minimum 
$$t_{\min}=\frac{\langle R_i-GS_i G^\top,G\d S_i G^\top\rangle}{\|G\d S_i G^\top\|^2}.$$
Therefore, new iterate is $(S_i)_{new}=\max(S_i+t_{\min}\d S_i,0).$

\section{Approaches based on search space transformation}
\label{sec:ELS_ADAM}

The third and the fourth method that we developed for the \ref{eqn:SNMTF} are based on so-called space transformation, where we replace the non-negativity constraints on $G$ and $S_i$ by 
transforming objective function SE and we then apply the gradient descent method with exact line search (GM-ELS) and ADAM (adaptive moment estimation) from \cite{adam}.
We can transform the feasible set for the \ref{eqn:SNMTF} $G$ (search space) from $\R^{n\times k}_+$ to $\R^{n\times k}$ by a simple variable substitution
\begin{align}
    G &= f(\tilde{G}) \\
    S_i &= f(\tilde{S}_i),
\end{align}
where $G$ is a non-negative matrix, $\tilde{G}$ is a real matrix and $f$ is an appropriate 
transformation function.
Similarly, we can transform the feasible sets for $S_i$ from $(\cS_k)_+$ to $(\cS_k)$ using the same transformation.

The transformed optimization problem, with $\tilde{G}, \tilde{S}_i$ as new variables, is unconstrained optimization problem, however, this gain is traded for a more complicated objective function
\begin{equation}
    \label{math:transformedSE}
    \SE = \sum_i \|R_i - f(\tilde{G}) f(\tilde{S}_i) f(\tilde{G})^\top\|^2.
\end{equation}
Advantage of this approach is that common local search algorithms developed for unconstrained optimization problems can be applied without modifications.

This technique has a long history in non-linear optimization and is usually employed in order to transform bounded search space variables to unbounded. Examples of transformations used in literature include smooth transformation of open hyper-rectangle to $\mathbb{R}^m$ \cite{transformation}, continuous transformation of closed hyper-rectangle to $\mathbb{R}^m$ \cite{trans2} and even continuous transformation of closed hyper-rectangle to search space with toroidal topology \cite{trans3}.

Two transformation functions  that we  consider in this work are:
\begin{itemize}
    \item Element-wise absolute value (search space boundary extension by mirroring \cite{trans2} adapted for non-negativity)
    \begin{equation}
        \label{math:f1abs}
        f_1(X) = \lvert X \rvert
    \end{equation}
    \item Element-wise square
    \begin{equation}
        f_2(X) = X \odot X
    \end{equation}
\end{itemize}
We can apply $f_1$ and $f_2$ to any matrix, regardless its dimension.
The benefit of using absolute value is that this transformation does not change characteristics of the optimization landscape in any way. 
However, by using absolute value, the objective function ceases to be a polynomial. Element-wise square, on the other hand, changes the optimization landscape, however, objective function remains a polynomial, but with doubled  degree.

\subsection{Gradient calculation}
\label{ch:gradcalc}

With this  optimization approach we have gradients or subgradients in every  feasible point, which can be computed using the chain rule. By introducing 
\begin{equation}\label{eq:Di}
    \D_i = R_i - f(\tilde{G}) f(\tilde{S}_i) f(\tilde{G})^\top,
\end{equation}
and by using that  $R_i$ and $S_i$ are symmetric and that $f$ does not break the symmetry, the gradient of $\SE$ can be expressed as 
\begin{align}
    \label{math:grad1}
   \d G ~=~ \grad_{\tilde{G}}\SE &= -4 \sum_i f'(\tilde{G})\odot\left(\D_i\, f(\tilde{G}) f(\tilde{S}_i)^\top \right) \\
    \label{math:grad2}
    \d S_i~=~\grad_{\tilde{S}_i}\SE &= -2 \sum_i f'(\tilde{S}_i)\odot\left(f(\tilde{G})^\top  \D_i\, f(\tilde{G})\right),
\end{align}
where $f'$ is the derivative of function $f$, applied component-wise, i.e. $f'(\tilde G) = (f'(G_{ij}))_{i,j}$. 
Formulae \eqref{math:grad1} and \eqref{math:grad2} are derived in appendix \ref{sec:grad}.

Function $f$ may also be only piecewise differentiable. This is the case for  $f_1(X)=\lvert X\rvert$ from above, which is non-diferentiable at any component of $X$ equal to $0$. 
In such a situation, we take in the non-diferentiable points the  sub-derivatives. In the case of $f_1$, the set of all sub-derivatives is  $[-1, 1]$. Therefore, we are free to make a choice $f_1'(0)=0$ which is also in accordance with the statement of Fermat's theorem on stationary points which says that every local extreme is a stationary point (the function derivative is zero at that point).

This is a very common approach in gradient based optimization, especially in deep learning. However, the proof of mathematical validity of using subgradients in stochastic  optimization is, despite its wide use, remarkably recent \cite{subgradient}. 

\subsection{Gradient method with exact line search (GM-ELS)}

In this subsection we present an algorithm employing the line search strategy to the transformed problem. This is an iterative procedure of making steps in direction of gradient of $\SE$ using
step size $t$ which is calculated independently for each specific step made so that the step taken minimizes $\SE$.
\begin{gather}
    \label{math:gupdate}
    \tilde{G} \assign \tilde{G} - t \, \d \tilde{G} \\
    \label{math:supdate}
    \tilde{S}_i \assign \tilde{S}_i - t \, \d \tilde{S}_i
    \\
    \label{math:alphaopt}
    t = \argmin_t \SE(\tilde{G}-t\,\d G,\; \tilde{S}_i-t\,\d \tilde{S}_i)
\end{gather}
Optimisation problem \eqref{math:alphaopt} needs to be solved in every iteration of the algorithm. Choice of function $f$ affects the form of this optimization problem. In general (eg.\ $f=f_1$), step size calculation \eqref{math:alphaopt} is a nonlinear optimization problem which can only be solved approximately using inexact methods. However, by choosing $f=f_2$ minimisation \eqref{math:alphaopt} becomes a uni-variate  minimisation of a polynomial of order 12 which can be solved exactly (up to numerical errors). We call the algorithm employing this technique gradient method with exact line search (GM-ELS).

By choosing $f=f_2$ the objective function becomes
\begin{equation}
    \label{math:SEpoly}
    \SE = \sum_i\big\lVert R_i-\tilde{G}^2\tilde{S}_i^2\tilde{G}^{2\top}\big\rVert^2,
\end{equation}
where squares act element-wise. To find out how SE changes when moving in direction of gradient we substitute $\tilde G$ and $\tilde{S}_i$ in  \eqref{math:SEpoly} with right hand sides of \eqref{math:gupdate} and \eqref{math:supdate}. We observe that SE changes with $t$ as a polynomial
\begin{equation}
    \label{math:poly}
    p(t) = 
    \sum_{i=0}^{12} c_i t^i.
\end{equation}
By finding $t$ that minimizes $p(t)$ we find what step size to make in direction of gradient so that SE is minimal. Even though minimisation of polynomial \eqref{math:poly} is an easy problem, 
calculation of its coefficients is computationally quite expensive. Initial tests show that the this calculation is approximately 7 times more costly compared to the calculation of the gradient. 
We also tried to reconstruct coefficients of $p(t)$ by polynomial curve fitting methods using it evaluations on $\ge 13$ points, but the step lengths obtained this way yielded worse performance. See our comment on page \pageref{note:polyfit} and Section  \ref{sec:poly} for details on calculation of these coefficients. Full GM-ELS algorithm is stated in Fig.~\ref{fig:lsalg}.

Due to exact calculation of optimum $t$, SE monotonically decreases during iterations, hence it converges to some non-negative value.
This means that at some iteration we have 
\begin{equation}
    \min p(t) \approx  p(0).
\end{equation}
The  equality can (approximately)  hold only if the polynomial $p(t)$ has a minimum at point $t=0$ which indicates that the gradient of $\SE$ is zero and proves convergence of GM-ELS toward a point with zero gradient.

\begin{figure}
\caption{Exact line search algorithm}
\label{fig:lsalg}
\fbox{\parbox{\textwidth}{
\begin{enumerate}
\item[]{\bf Input:} Non-negative symmetric $R_1,\ldots,R_N$.
\item[1.] {\bf Initialisation:} Compute initial  values for $\tilde{G}$ and $\tilde{S}_i$.
\item[2.]{\bf While} termination test not satisfied 
\begin{itemize}
    \item [2.1]	Compute gradients $\d G=\grad_{\tilde{G}}\SE$ and $\d S_i=\grad_{\tilde{S}_i}\SE$ using \eqref{math:grad1} and \eqref{math:grad2} with ${f=f_2}$.
    \item [2.2]	Compute coefficients of polynomial $p(t)$ using \eqref{math:coeff}.
    \item [2.3]	$t \assign \argmin p(t)$
    \item [2.4]{\bf For} $X\in\{\tilde{G}, \tilde{S}_1, \ldots, \tilde{S}_N\}$
    \begin{itemize}
        \item [2.4.1] $X \assign X - t \, \grad_{X}\SE$.
    \end{itemize}
\end{itemize}
\item[3.]{\bf Return} $\tilde{G} \odot \tilde{G},\ \tilde{S}_1 \odot \tilde{S}_1,\ldots,\tilde{S}_N \odot \tilde{S}_N$
\end{enumerate}
}}
\end{figure}

\subsection{Adaptive moment estimation}

The adaptive moment estimation (ADAM) \cite{adam} is a gradient based optimization algorithm and is currently state-of-the-art algorithm for training deep neural networks. It was designed for problems with high dimensional search space, noisy objective function and sparse gradients. The ADAM cannot be straightforwardly applied for solving the \ref{eqn:SNMTF} due to non-negativity constraint. However, this limitation can be circumvented by using transformed formulation of the \ref{eqn:SNMTF} presented in \eqref{math:transformedSE} which releases the non-negativity constraint.

In order to apply ADAM to the \ref{eqn:SNMTF} we chose to transform search space by using element-wise absolute value, ie.\ ${f=f_1}$. We decided to use absolute value because this operation is the most computationally efficient way to guaranty non-negativity, since it can be implemented with a conditional clause and no other operations. Also, this transformation leaves optimization landscape unchanged. The use of absolute value imply the use of sub-derivatives as explained in section \ref{ch:gradcalc}.

In ADAM algorithm steps are not made in direction of gradient but in direction of moving average of the gradient. Another trait is step size scaling which is adapted for each matrix element independently. Update rule for both $G$ and $S_i$ are the same, therefore we only state the one for $G$, which is updated using the following formulae:

\begin{align}
    \label{math:momentum}
    M_{\tilde{G}} &\assign \beta_1 M_{\tilde{G}}+(1-\beta_1)\, \d \tilde{G} \\
    \label{math:secmomentum}
    V_{\tilde{G}} &\assign \beta_2 V_{\tilde{G}} + (1 - \beta_2)\, \d \tilde{G}\odot\d \tilde{G} \\
    \label{math:adamupdate}
    \tilde{G} &\assign \tilde{G}-\eta M_{\tilde{G}}  \oslash (\sqrt{V_{\tilde{G}}}+\varepsilon),
\end{align}
where $\oslash$ denotes element-wise division of two matrices and $\sqrt{\phantom{a}}$ and adding $\varepsilon$ are   applied element-wise. The  Greek symbols are scalars and $M_{\tilde{G}}$ and $V_{\tilde{G}}$ are the matrices of the same size as $\tilde{G}$ and contain the information about the mean and the  variance of gradient  $\d G$, respectively. Parameter $\eta$ changes with a predetermined schedule as stated in Fig.\ \ref{alg:adam}.

As seen in  (\ref{math:momentum}), ADAM uses momentum $M$ which is an exponential moving average of the gradient with $\beta_1$ being the decay constant. Using momentum instead of gradient for descent direction makes optimization more stable by preventing oscillations and by averaging possibly noisy or sparse gradients~\cite{momentum}. The ADAM also uses second moment $V$ \eqref{math:secmomentum} which is an exponential moving average of element-wise square of the gradients. This quantity is an approximation of diagonal of Fisher information matrix and is used to adaptively scale the step size for each search variable independently as seen in  \eqref{math:adamupdate}.

The complete algorithm for solving the \ref{eqn:SNMTF} is shown in Fig.~\ref{alg:adam} and is controlled by four parameters ($\alpha$, $\beta_1$, $\beta_2$ and $\varepsilon$) which influence its efficiency. The most important one is the so called learning rate $\alpha$. The usual magnitude of the learning rate is $\alpha\sim10^{-3}$, however, due to descent direction scaling by using second moment $V$, algorithm efficiency is intended to be robust with respect to this choice. Another important parameter is $\beta_1$ which should be chosen in accordance to the noise level and gradient sparsity level. However, there are no theoretically founded ways to assess appropriate parameter values based on the optimization problem at hand. Therefore some degree of parameter tuning is normally used.

Empirically, ADAM has proven to be very efficient and with a superior convergence rate compared to other gradient based optimization algorithms \cite{adam}. However, contrary to these findings there are no guaranties that the algorithm converges. In \cite{adamconvergence} explicit example of a simple convex optimization problem was provided where ADAM does not converge to the optimal solution. In practice, however, such convergence problems are seldomly encountered and can be mended by changing initial point and/or algorithm parameters which influence algorithm behaviour.

\begin{figure}
\caption{ADAM algorithm}
\label{alg:adam}
\fbox{\parbox{\textwidth}{
\begin{enumerate}
\item[] {\bf Input:} Non-negative symmetric $R_1,\ldots,R_N$ \\ \phantom{{\bf Input:}} and parameters $\alpha, \beta_1, \beta_2, \varepsilon$.
\item[1.] {\bf Initialisation:} Compute initial values for $\tilde{G}$ and $\tilde{S}_i$.
    \item [2.] {\bf For} $X\in\{\tilde{G}, \tilde{S}_1, \ldots, \tilde{S_N}\}$
    \begin{itemize}
        \item [2.1] $M_X \assign X\cdot0$
        \item [2.2] $V_X \assign X\cdot0$
    \end{itemize}
    \item [3.] $i \assign 0$
\item[4.]{\bf While} termination test not satisfied 
\begin{itemize}
    \item [4.1] $i \assign i+1$
    \item [4.2] $\eta \assign \alpha\frac{\sqrt{1-(1-\beta_2)^i}}{1-(1-\beta_1)^i}$
    \item [4.3]	Compute derivatives $\d \tilde G$,  $\grad_{\tilde{S}_i}\SE,\ldots, \grad_{\tilde{S}_N}\SE$ using \eqref{math:grad1} and \eqref{math:grad2} with $f=f_1$.
    \item [4.4] {\bf For} $X\in\{\tilde{G}, \tilde{S}_1, \ldots, \tilde{S}_N\}$
    \begin{itemize}
        \item [4.4.2]	$M_X \assign \beta_1M_X+(1-\beta_1)\, \grad_{X}\SE$
        \item [4.4.3] $V_X \assign \beta_2V_X + (1 - \beta_2)\, \grad_{X}\SE\odot\grad_{X}\SE$
        \item [4.4.4] $\displaystyle X \assign X - \eta  M_X\oslash (\sqrt{V_X}+\varepsilon)$
    \end{itemize}
\end{itemize}
\item[5.]{\bf Return} $|\tilde{G}|,\ |\tilde{S}_1|,\ldots,|\tilde{S}_N|$
\end{enumerate}
}}
\end{figure}

\section{Implementation details}
\label{sec:implementation}

In this section we explain the details about  implementation of all four algorithms described in the previous sections. The FPM and BCD were implemented in Matlab \cite{MATLAB:2019}. The reason to choose this framework was that the development of this code was part of another project \cite{malod2019towards} which imposed Matlab as working environment. 
The ADAM and GM-ELS were implemented later with ambition to be high-performance open-source code. This is why we used  Python and TensorFlow library \cite{tf}.
The Matlab and Python codes and all the data-sets used in this paper are available on GitLab \cite{git}. 

\subsection{Starting points }\label{sec:starting_points}
During our study, we have observed a well-known fact that good starting points are often very important for overall convergence. By default, we start the methods using random starting matrices. More precisely, for FPM we generate $G$ in Matlab by a simple call {\tt rand(n,k)}, while each matrix $S_i$ is generated by {\tt rand(k)} and symmetrisation afterwards. Likewise, we generate starting $G$ for BCD, while the first tuple of $S_i$ is computed from the starting $G$. 
For GM-ELS and ADAM, we generate staring points for both $G$ and $S_i$ by using NumPy's function {\tt random.rand}.

Following the literature \cite{boutsidis2008svd}, we also tested deterministic starting points $G$, extracted from 
eigenvalue decomposition of (symmetric) matrix $R=\sum_i R_i$. We computed for the given inner dimension $k$ the $k$ largest magnitude eigenvalues
$|\lambda_1|\ge |\lambda_2|\ge \cdots \ge |\lambda_{k}| $ and the corresponding eigenvectors $\x_i$. For each eigenvector $\x_i$ we considered its positive and negative parts, $\x_i^+$ and $\x_i^-$, respectively, and took the one with larger euclidean norm (we denote it by $\tilde\x_i$). By concatenating the $k$ vectors $\tilde\x_i$  column-wise we finally obtain the $n\times k$ starting point $G$. Spectral decomposition was performed using function {\tt eigs} available in Matlab or Scipy's module {\tt sparse.linalg}. Note that we could also compute eigenvalues using function {\tt svds} but our tests showed that {\tt eigs} was always faster, which  is not surprising.


By extensive tests where we compared for each instance the best $\MSE$ and the arithmetic mean value of the $\MSE$s over 30 runs with different random starting matrices with the $\MSE$ obtained by deterministic starting point we noticed that arithmetic mean values were almost always worse than the deterministic $\MSE$. By using $t$-tests we observed that the differences were usually statistically significant, but in practice these differences were  small (below $0.01$ -- the significance came from the very small variances of the $\MSE$s).
On the other hand, the best (minimum) value of $\MSE$ out of 30 values obtained with different random starts was always very close to the $\MSE$ obtained by deterministic start, several times slightly better.
This suggests that if the dimension of the problem is small then making several rounds with different random starting points and taking the best value is reasonable, but for higher dimensions starting with the deterministic $G$ is the best strategy.

All our computations were  done with deterministic starting points only, since all three data sets included also large scale matrices.

\subsection{The main parts of the algorithms}

After the initialisation of FPM and BCD, we iterate through the main parts of the algorithms.
For FPM this means that we iteratively compute $(S_1,\ldots,S_N)_{new}$ using \eqref{eqn:MU2a} and $G_{new}$
using \eqref{eqn:MU1a} after adding Matlab's defined constant ${\tt eps} = 2.2204\cdot 10^{-16}$ to all of the denominators.

For BCD, the updates $(S_1,\ldots,S_N)_{new}$ and $G_{new}$ (Steps 2.1 and 2.3 from Figure \ref{fig:1})
are computed using 10 iterations of projected gradient method with exact line search. In both cases, in every iteration, we compute gradient at the current point and the new iterate is obtained by exact line search as described in Section \ref{sec:BCD_PGM}.
Finally, we project each new iterate to the cone of non-negative matrices.

The GM-ELS and ADAM algorithms were implemented using TensorFlow library. The major advantages of this framework are support for automatic differentiation and highly efficient compiler which results in optimized computational graph for gradient calculation with built-in parallelisation.


A single iteration of GM-ELS starts with gradient calculation. This step is implemented using TensorFlow functions for reverse automatic differentiation. In other words, computational graph for SE calculation is recorded and differentiated using chain rule and dynamic programming methods at compile time, yielding  an optimized computational graph for calculation of gradient of SE 
with respect to both $\tilde{G}$ and $\tilde{S}_i$. The resulting gradient is equal 
to the one stated in  \eqref{math:grad1} and \eqref{math:grad2}. We chose this method because it is computationally efficient and requires less programming effort and is therefore less error prone. Knowing the gradient, a polynomial is calculated which tells how $\SE$ changes with step size in direction of gradient. 
This step is
computationally intensive and is
covered in detail in appendix \ref{sec:poly}.
In our case these computations are left to TensorFlow library which manages to do them  efficiently due to its compiler optimizations such as elimination of common subexpressions and simplification of arithmetic statements. However, the time needed for this step was still non-neglectable (approximately 7 times  longer compared to gradient calculation). 
 In the next step polynomial is minimized  numerically to find the step size that minimizes SE. Minimization was performed by finding all roots of derivative of this polynomial. We used NumPy's function {\tt roots} that relies on computing the eigenvalues of the companion matrix. Step that minimizes $\SE$ is finally employed to update the matrices $G$ and $S_i$. 

The ADAM algorithm makes use of the same method of gradient calculation as GM-ELS. Because of search space transformation there is no non-negativity constraint, therefore, standard implementation of ADAM was used as implemented in TensorFlows module {\tt keras.optimizers.Adam}. Unlike other three algorithms, ADAM also requires optimization parameters to be set by the user. We used parameter tuning to find appropriate parameter values as explained in Section \ref{sec:tuning}.

\subsection{Termination criterion}\label{subsect:term_criteria}

The stopping criteria for all methods were maximum number of iterations or if the $\MSE$ falls below $10^{-2}$ or if absolute difference between the $\MSE$ in given iteration and the $\MSE$ in the
 previous iteration is below $10^{-10}$. We did an extensive testing on small, medium and large scale matrices, see Section \ref{subsec:results}, and finally experimentally set the maximum number of iterations to  4000 for FPM, 300 for BCD, 1000 for GM-ELS and 3000 for ADAM. With these values the computation times (wall time) of all four methods were of the same order.
 A special section (Section \ref{sect:mse_vs_time}) is devoted to the analysis how $\MSE$ decreases with time.

\subsection{Parameter tuning}
\label{sec:tuning}
Behaviour of ADAM algorithm is controlled by four parameters ($\alpha, \beta_1, \beta_2, \varepsilon$). To find suitable values we performed parameter tuning on $\alpha, \beta_1, \beta_2$ while leaving $\varepsilon$ fixed since it is the least important parameter which only prevents division by zero \cite{adam}. Tuning was performed using random search which is more efficient compared to commonly used grid search \cite{partuning}. We randomly generated 100 points in parameter space on a hyper-rectangle defined by $\alpha\in[10^{-4},10^{-1}]$, $\beta_1\in[0.2, 0.999]$ and $\beta_2\in[0.1, 0.999]$. 
Each of the 100 parameter combinations was evaluated by performing three runs of ADAM algorithm on each problem included in a synthetic data-set (the first data-set described in Section \ref{sec:data}) with $k=K$, starting from a random initial point.
By construction we know that for each 5-tuple of matrices $(R_1,\ldots,R_5)$ from this data-set the optimum $\MSE$ is 0.

\begin{figure}
    \centering
    \begin{tabular}{cccc}
        \includegraphics[height=0.35\textwidth]{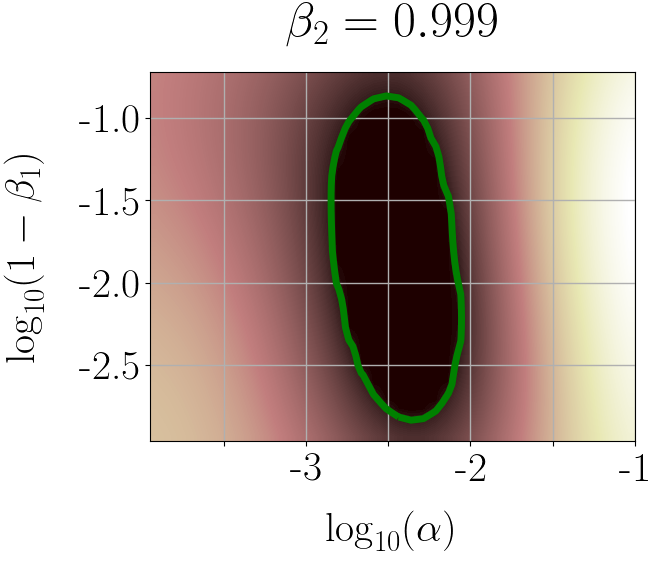} &
        \includegraphics[height=0.35\textwidth]{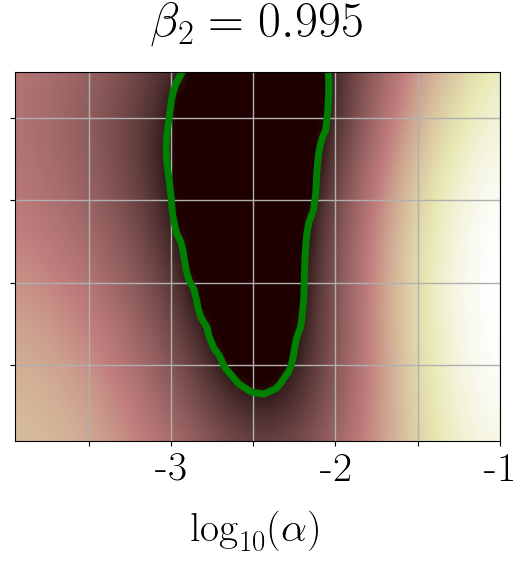} &
        \includegraphics[height=0.35\textwidth]{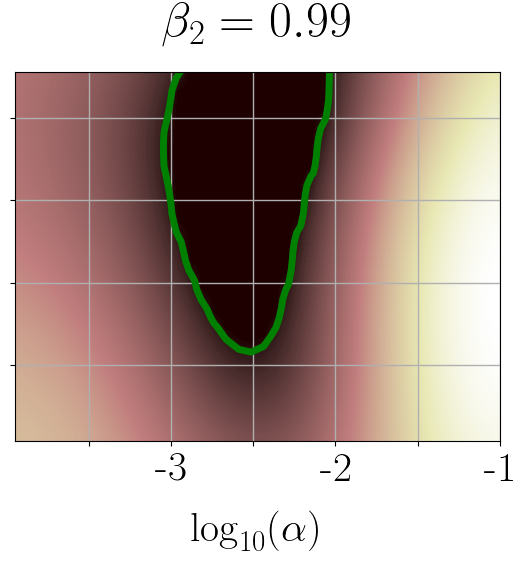} &
        \includegraphics[height=0.32\textwidth, trim={13.2cm -2.4cm 0 0}, clip]{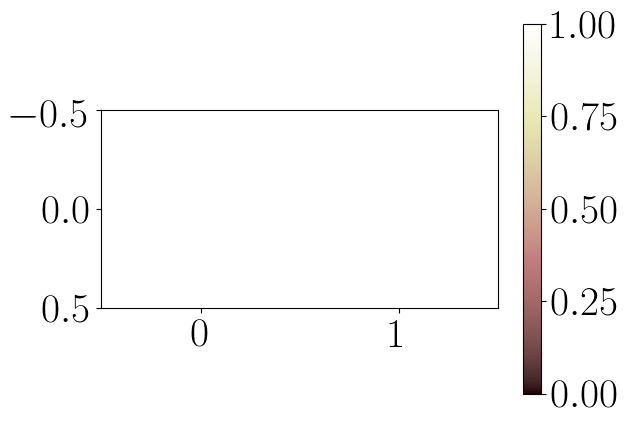} \\
        \includegraphics[height=0.35\textwidth]{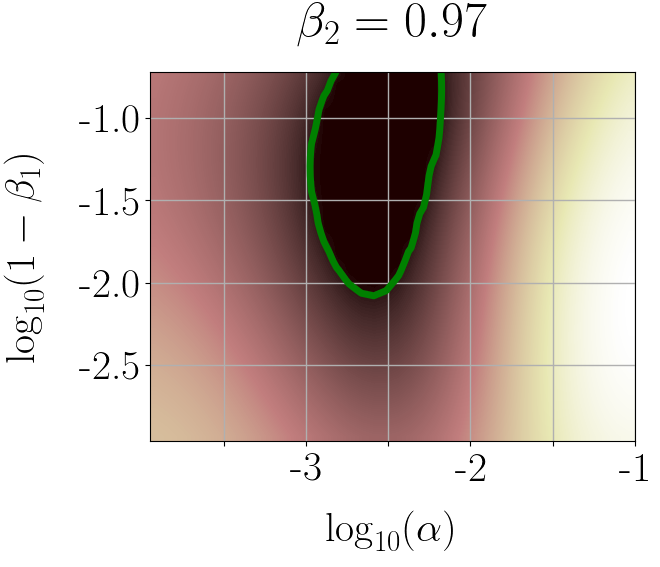} &
        \includegraphics[height=0.35\textwidth]{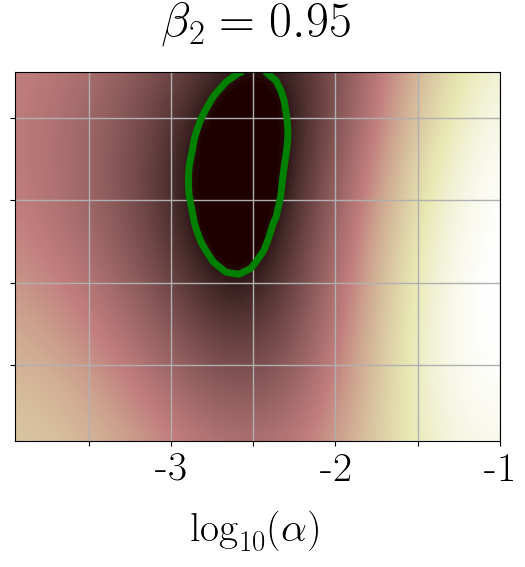} &
        \includegraphics[height=0.35\textwidth]{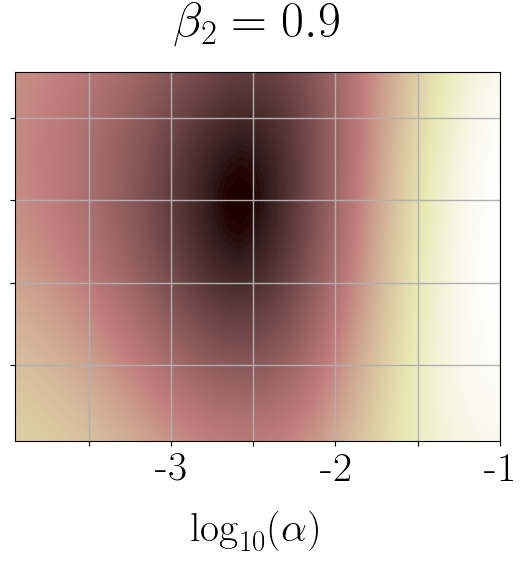} &
        \includegraphics[height=0.32\textwidth, trim={13.2cm -2.4cm 0 0}, clip]{images/adam_pars6.png} \\
    \end{tabular}
    \caption{Impact of parameters $\alpha, \beta_1$ and $\beta_2$ on efficiency of ADAM algorithm for
    synthetic problems with $K=k$. Heatmaps show maximum
    (over problems) mean (over runs) MSE with respect to $\alpha$ and $\beta_1$ for different
    values of $\beta_2$. Green contours envelop areas where $\MSE \leq 0.01$ for all problems in synthetic data-set. Visualisation
    was made using support vector regression model trained on a set of 100 randomly chosen triples
    of parameters $(\alpha, \beta_1, \beta_2)$.}
    \label{fig:adampars}
\end{figure}

We observe that matrix size $n$ and latent dimension $k$ has little influence on the position of optimal parameter values. In fact the $\MSE$ is quite low on a large proportion of inspected parameter space. Fig.~\ref{fig:adampars} shows a region of parameter space where the $\MSE \leq 0.01$ for all problems from the synthetic data-set and represents a guideline for choosing appropriate parameters of ADAM algorithm for this class of optimization problems. Even though there is no guaranty that this region of parameter space will be promising for other, real-world, factorization problems, we conclude from these results that small changes in the parameters do not considerably change the algorithm efficiency. This means that good results can be acquired even with minimal tuning.

Based on these findings we fixed ADAM parameters to $\alpha=0.002$, $\beta_1=0.95$, $\beta_2=0.995$ and $\varepsilon=10^{-8}$. All reported results for ADAM in the following sections were acquired using these parameter values.

\renewcommand{\tabcolsep}{0pt}
\renewcommand{\tabcolsep}{6pt}

\section{Results}
\label{sec:results}

\subsection{Data}
\label{sec:data}

We benchmark the four methods presented in the previous sections on three data-sets: on a set of synthetic random
matrices,
real-world similarity matrices gathered by \cite{realdata} and biological network data gathered during our previous studies. 
All three data-sets are included in the afore  mentioned GitLab repository \cite{git}.

The set of synthetic random matrices was constructed by us. It consists of 5-tuples of non-negative symmetric matrices of order $n=100, 200, 500, 1000, 2000$ and $5000$. For each $n$ we first
select $K\in \{10,20,30,40,50\}$ and construct random non-negative $G$ of order $n\times K$ with orthogonal columns 
and 5 non-negative matrices $S_i$ of order $K\times K$ with densities approximately 65\% and computed matrices $R_i$ as products $R_i=GS_iG^\top$. For each possible $n, K$ we store matrices $R_i, S_i$ and $G$ into a file and make them available, e.g., directory {\tt test\_data\_n=1000\_k=50\_N=5} contains matrices $R_1,R_2,\ldots,R_5, G, S_1,\ldots,S_5$ corresponding to $n=1000,K=50$.

The second data-set was gathered by \cite{realdata} and represents real-world similarity or proximity matrices from different fields of data science. Out of 13 available data-sets, only 5 of them have only non-negative values:
\begin{itemize}
\item \textbf{Aural} data-set is taken from \cite{4098942}, investigating the human ability to distinguish different types of sonar signals by ear. The signals were returns from a broadband active sonar system, with 50 target-of-interest signals and 50 clutter signals. Every pair of signals was assigned a similarity score from 1 to 5 by two randomly chosen human subjects unaware of the true labels, and these scores were added to produce a $100 \times 100$ similarity matrix with integer values from 2 to 10.

\item\textbf{Protein} data-set has sequence-alignment similarities for 213 proteins from 4 classes, where classes one through four contain 72, 72, 39, and 30 points, respectively \cite{DBLP:journals/pami/HofmannB97}.

\item\textbf{Voting} data-set comes from the UCI Repository. It is a two-class classification problem with 435 points, where each sample is a categorical feature vector with 16 components and three possibilities for each component. Value difference metric was computed \cite{Stanfill:1986:TMR:7902.7906} from the categorical data, which is a dissimilarity that uses the training class labels to weight different components differently so as to achieve maximum probability of class separation.

\item\textbf{FaceRec} data-set consists of 945 sample faces of 139 people from the NIST Face Recognition Grand Challenge data-set. There are 139 classes, one for each person. Similarities for pairs of the original three-dimensional face data were computed as the cosine similarity between integral invariant signatures based on surface curves of the face \cite{4301238}.

\item\textbf{Zongker} data-set is a digit dissimilarity data (2000 points in 10 classes) and is based on deformable template matching. The dissimilarity measure was computed among 2000 handwritten NIST digits in 10 classes, with 200 entries each, as a result of an iterative optimization of the non-linear deformation of the grid \cite{zongker}.
\end{itemize}

The third data-set is biological molecular network data. 
It  consists of 4 molecular interaction networks 
(Protein-Protein Interaction network - PPI, Gene Co-expression COEX, Genetic Interaction - GI and sequence similarity networks - SeqSim)  for 5 different species  listed in Table~\ref{tab:speciesnames}. 
These networks  represent real-world use cases where several matrices
need to be factored simultaneously in order to integrate multiple data sources and
produce more consistent, accurate and useful information.  
The data was collected  in June 2016 during the preliminary phase of research published in \cite{icell} using BioGRID \cite{biogrid}, COXPRESdb \cite{coxpresdb},
STRING \cite{string} and BLAST \cite{blast}, respectively, and is available on our GitLab profile \cite{git}. 

\begin{table}
\begin{tabular}{lllr}
\hline
common name & taxonomic name & tag & $n$ \\
\hline
fission yeast  & Schizosaccharomyces pombe & Yeast\_SPO & 4979 \\
baker's yeast & Saccharomyces cerevisiae & Yeast\_SCE & 5890 \\
fruit fly & Drosophila melanogaster & Fly\_DME & 13190 \\
nematode worm & Caenorhabditis elegans & Worm\_CEL & 17519 \\
human & Homo sapiens & Human\_HSA & 20967 \\
\hline
\end{tabular}
\caption{Species considered in the biological network data-set.}
\label{tab:speciesnames}
\end{table}

\subsection{Results for $\MSE$ on the test data-sets}\label{subsec:results}

We run all four methods on all three data-sets using high-performance cluster at University of Ljubljana, Faculty of mechanical engineering, which is a E5-2680 V3 (1008 hyper-cores) DP cluster, with IB QDR interconnection, 164 TB of LUSTRE storage, 4.6 TB RAM, supplemented by GPU accelerators.

\begin{landscape}
\begin{figure}
    \centering
    \includegraphics[width=\linewidth, trim={4.8cm 2.2cm 4.8cm 3cm}, clip]{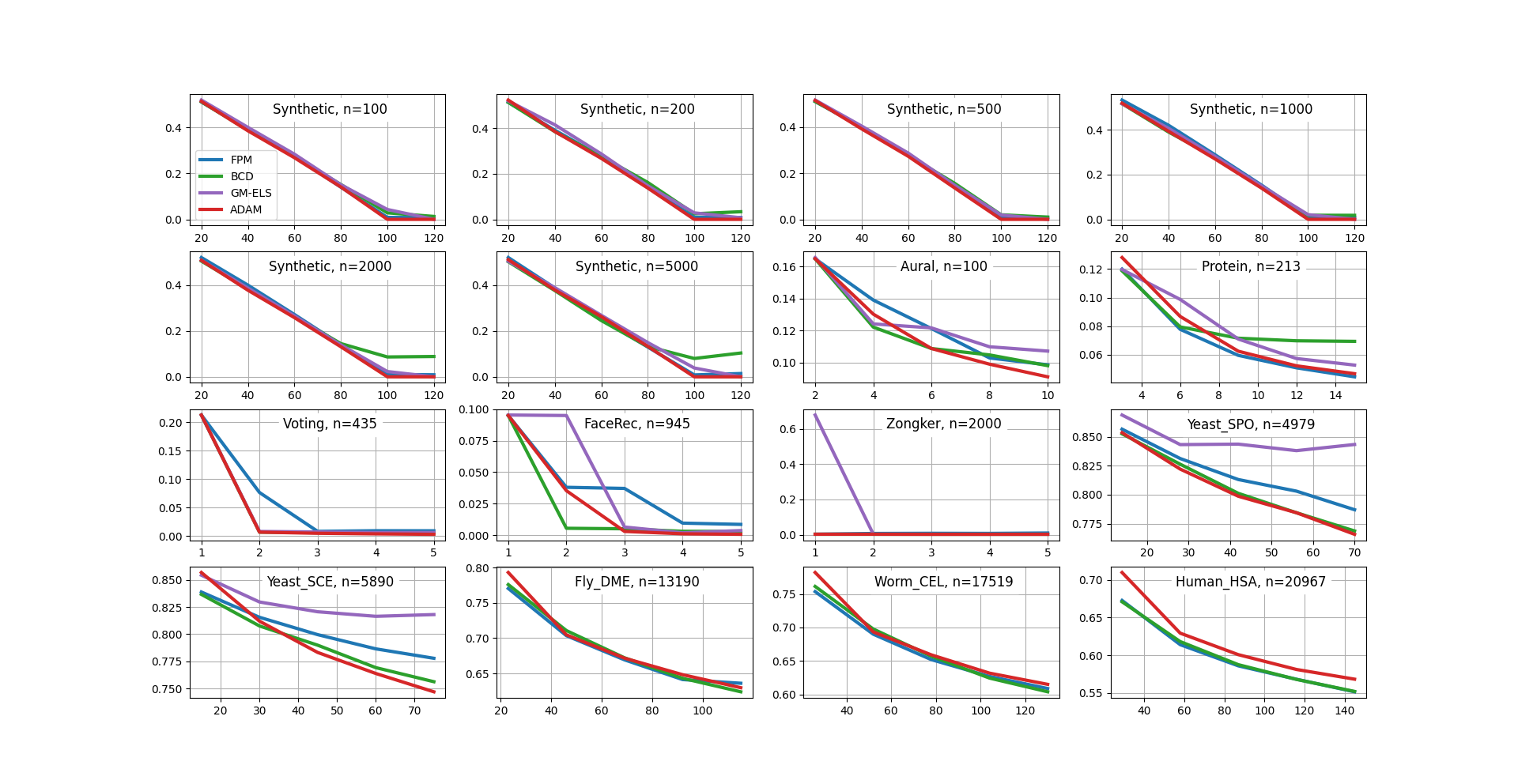}
    \caption{$\MSE$ with respect to inner  dimension $k$ for different algorithms and problems. It is a visual depiction of tables \ref{tab:synthetic}, \ref{tab:other} and \ref{tab:species}. In case of the synthetic problems, $k$ is measured in  \% of the actual latent dimension $K$ (as in Table \ref{tab:synthetic}).
    }
    \label{fig:MSE_k}
\end{figure}
\end{landscape}

Results obtained by all four methods on the data-set of synthetic random orthogonal matrices are reported in a condensed format in Table \ref{tab:synthetic} and in the first 6 plots of Fig.~\ref{fig:MSE_k}. 
For each $n\in\{100$, $200$, $500$, $1000$, $2000$, $5000\}$ and $K\in\{10$, $20$, $30$, $40$, $50\}$ we run all four methods with deterministic starting solution $G$ and with inner dimension equal to $k=0.2K$, $0.4K$, $0.6K$, $0.8K$, $1.0K$, $1.2K$. Procedure of choosing various $k$ values simulates real-world case where inner dimension $K$ is usually not known. However, in this case we know that for $k=1.0K,1.2K$, by construction, the optimum $\MSE_{opt}=0$ and our results demonstrate the capability of the methods we used to reach a global optimum.

Each row of Table \ref{tab:synthetic} shows the mean values of the $\MSE$, for each $n$ and each possible value of $k/K$.
For example, the last row of this table shows the mean values of $\MSE$ for $n=5000$ and for $k/K=120~\%$. This means that we computed and reported  for each algorithm the mean values of five $\MSE$s: for $k=12,K=10;~k=24,K=20;~k=36,K=30;~k=48,K=40$ and $k=60,K=50$.

Numerical results from Table \ref{tab:synthetic} reveal that FPM, GM-ELS and ADAM are very competitive regarding the $\MSE$, while BCD yields worse $\MSE$ for larger values of $n$ and $k/K$.
A deeper analysis shows that worse performance of BCD is due to small number of outer iterations (it was limited to 300). When we allow 1000 iterations, the resulting $\MSE$ was of the same order of magnitude -- i.e., below $10^{-2}$ as for the other three methods.

Therefore, based on the $\MSE$ and this first data-set we cannot decide which method is better. However, when we take into account also the time, situation changes -- see Section \ref{sect:mse_vs_time}

Table \ref{tab:other} contains results on the second data-set (real-world similarity matrices).
The four methods were run  on each data-matrix, using deterministic starting $G$. The inner dimensions were chosen such that they increase evenly towards the number of the clusters in the underlying data points. 
We can see that the mean values of $\MSE$ for {\tt Voting, FaceRec}, and {\tt Zonker} are falling rapidly towards 0 which demonstrate that all four methods are capable to find the optimum decompositions when $k$ approached the estimated real value of $K$ .
However, for {\tt Aural} and {\tt Protein} the $\MSE$ does not come very close to 0. This can mean that those two matrices cannot be factored in the proposed way  due to the noise or the inner dimension was not correctly identified or  some other hidden reason might be present.

For this data-set we observe no clear pattern that would declare which algorithms are systematically better than the others. For large $k$ all algorithms obtain similar MSE values (with exception of {\tt Protein} where BCD is clearly not as good as the others). In the case of smaller $k$ the ranking of the algorithms changes with respect to problem instance and inner dimension $k$.

Table \ref{tab:species} contains results of all four methods evaluated on the third data-set, which consists of 5 4-tuples of symmetric non-negative matrices with sizes varying from 4979 (Yeast\_SPO) to 20967 (Human\_HSA). All methods started in the deterministic starting point, as described in Section \ref{sec:starting_points}.

Solving \eqref{eqn:SNMTF} for this data was very time and memory consuming. We can see that GM-ELS runs out of memory (OOM) for the largest three tuples of matrices, while the other three methods always terminate regularly, according to stopping criteria described in Section \ref{subsect:term_criteria}.
We decided to evaluate methods with 5 inner dimensions:
$\round{\sqrt{n}/5},2\round{\sqrt{n}/5},\ldots,5\round{\sqrt{n}/5}$, where $\round{\cdot}$ denotes rounding to the nearest integer.

Regarding the $\MSE$ we observe that FPM, BCD, and ADAM perform similarly. For the largest instance (Human\_HSA) ADAM returns slightly worse $\MSE$, but the difference is quite small. Similar situation can be noticed in case of Yeast\_SPO and Yeast\_SCE, where MSEs for FPM algorithm are systematically slightly higher. GM-ELS, on the other hand, is consistently worse than others or, for larger data-sets, even runs out of memory. This indicates that GM-ELS is not an appropriate choice when factoring large data-sets. This is a result of considerable resources needed for calculation of 12 order polynomial used for determining ideal step size for GM-ELS.

\begin{table}[]
\footnotesize
\centering
\begin{tabular}{rrrrrr}
\hline
 $n$ & $k/K (\%)$ &    FPM &    BCD &    GM-ELS &   ADAM \\
\hline
  100 & 20 & 0.5112 & 0.5119    & 0.5196 & 0.5138 \\
  100 & 40 & 0.3924 & 0.3883    & 0.4008 & 0.3858 \\
  100 & 60 & 0.2749 & 0.2726    & 0.2841 & 0.2685 \\
  100 & 80 & 0.1402 & 0.1519    & 0.1506 & 0.1409 \\
  100 & 100 & 0.0085 & 0.0283    & 0.0430 & 0.0000 \\
  100 & 120 & 0.0079 & 0.0127    & 0.0001 & 0.0000 \\
\hline
 200 & 20 & 0.5136 & 0.5135  & 0.5197 & 0.5236 \\
 200 & 40 & 0.3916 & 0.3852  & 0.4151 & 0.3856 \\
 200 & 60 & 0.2759 & 0.2763  & 0.2873 & 0.2676 \\
 200 & 80 & 0.1389 & 0.1619  & 0.1473 & 0.1375 \\
 200 & 100 & 0.0092 & 0.0246 & 0.0286 & 0.0000 \\
 200 & 120 & 0.0085 & 0.0335 & 0.0070 & 0.0000 \\
\hline
  500 & 20 & 0.5109 & 0.5104  & 0.5169 & 0.5151 \\
  500 & 40 & 0.3986 & 0.4015  & 0.4042 & 0.3919 \\
  500 & 60 & 0.2789 & 0.2768  & 0.2873 & 0.2733 \\
  500 & 80 & 0.1420 & 0.1561  & 0.1479 & 0.1360 \\
  500 & 100 & 0.0084 & 0.0199 & 0.0189 & 0.0000 \\
  500 & 120 & 0.0090 & 0.0096 & 0.0001 & 0.0000 \\
\hline
1000 & 20 & 0.5323 & 0.5159      & 0.5154 & 0.5183 \\
1000 & 40 & 0.4197 & 0.3895      & 0.4059 & 0.3928 \\
1000 & 60 & 0.2881 & 0.2777      & 0.2810 & 0.2694 \\
1000 & 80 & 0.1528 & 0.1444      & 0.1489 & 0.1393 \\
1000 & 100 & 0.0107 & 0.0183     & 0.0210 & 0.0000 \\
1000 & 120 & 0.0084 & 0.0180     & 0.0002 & 0.0000 \\
\hline
2000 & 20 & 0.5192 & 0.5036    & 0.5072 & 0.5071 \\
2000 & 40 & 0.3992 & 0.3830    & 0.3864 & 0.3770 \\
2000 & 60 & 0.2714 & 0.2614    & 0.2655 & 0.2578 \\
2000 & 80 & 0.1401 & 0.1447    & 0.1391 & 0.1308 \\
2000 & 100 & 0.0103 & 0.0865   & 0.0231 & 0.0001 \\
2000 & 120 & 0.0090 & 0.0885   & 0.0001 & 0.0000 \\
\hline
5000 & 20 & 0.5042 & 0.5021    & 0.5042 & 0.5114 \\
5000 & 40 & 0.3847 & 0.3823   & 0.3892 & 0.3790 \\
5000 & 60 & 0.2644 & 0.2704   & 0.2685 & 0.2627 \\
5000 & 80 & 0.1383 & 0.1446   & 0.1497 & 0.1329 \\
5000 & 100 & 0.0112 & 0.0415 & 0.0387 & 0.0001 \\
5000 & 120 & 0.0130 & 0.0739 & 0.0004 & 0.0001 \\ 
\hline
\end{tabular}
\caption{In this table we report $\MSE$ values obtained by all four algorithms on the synthetic data-set.
We report mean values of $\MSE$ taken over all pairs $(k,K)$ with ratio $100k/K=20,40,\ldots,120$}
\label{tab:synthetic}
\end{table}

\begin{table}[h]
\centering
\begin{tabular}{lrrrrrr}
\hline
instance & $n$ & $k$ & FPM & BCD & GM-ELS & ADAM \\
\hline
Aural   &  100 &  2 & 0.1649 & 0.1649  & 0.1657 & 0.1649 \\
Aural   &  100 &  4 & 0.1391 & 0.1222  & 0.1242 & 0.1303 \\
Aural   &  100 &  6 & 0.1213 & 0.1088  & 0.1217 & 0.1088 \\
Aural   &  100 &  8 & 0.1029 & 0.1048  & 0.1099 & 0.0990 \\
Aural   &  100 & 10 &  0.0986 & 0.0981 & 0.1072 & 0.0911 \\
\hline
Protein &  213 &  3 & 0.1199 & 0.1187  & 0.1196 & 0.1280 \\
Protein &  213 &  6 & 0.0778 & 0.0795  & 0.0987 & 0.0868 \\
Protein &  213 &  9 & 0.0596 & 0.0716  & 0.0710 & 0.0625 \\
Protein &  213 & 12 &  0.0509 & 0.0698 & 0.0574 & 0.0523 \\
Protein &  213 & 15 &  0.0446 & 0.0694 & 0.0528 & 0.0468 \\
\hline
Voting  &  435 &  1 &0.2129 & 0.2129 & 0.2129 & 0.2129 \\
Voting  &  435 &  2 &0.0766 & 0.0069 & 0.0081 & 0.0069 \\
Voting  &  435 &  3 &0.0083 & 0.0059 & 0.0073 & 0.0049 \\
Voting  &  435 &  4 &0.0094 & 0.0052 & 0.0062 & 0.0039 \\
Voting  &  435 &  5 &0.0093 & 0.0053 & 0.0063 & 0.0031 \\
\hline
FaceRec &  945 &  1 & 0.0951 & 0.0951 & 0.0955 & 0.0951 \\
FaceRec &  945 &  2 & 0.0380 & 0.0054 & 0.0951 & 0.0353 \\
FaceRec &  945 &  3 & 0.0371 & 0.0050 & 0.0064 & 0.0029 \\
FaceRec &  945 &  4 & 0.0095 & 0.0030 & 0.0018 & 0.0010 \\
FaceRec &  945 &  5 & 0.0085 & 0.0029 & 0.0036 & 0.0006 \\
\hline
Zongker & 2000 & 1 & 0.0025 & 0.0025 & 0.6779 & 0.0025 \\
Zongker & 2000 & 2 & 0.0068 & 0.0025 & 0.0025 & 0.0025 \\
Zongker & 2000 & 3 & 0.0077 & 0.0025 & 0.0024 & 0.0020 \\
Zongker & 2000 & 4 & 0.0072 & 0.0025 & 0.0023 & 0.0020 \\
Zongker & 2000 & 5 & 0.0094 & 0.0025 & 0.0024 & 0.0020 \\
\hline
\end{tabular}
\caption{MSE values obtained by all four algorithms on the real-world data-sets gathered by \cite{realdata}.}
\label{tab:other}
\end{table}


\begin{table}[h]
\centering
\begin{tabular}{lrrrrrr}
\hline
instance & $n$ & $k$ & FPM & BCD & GM-ELS & ADAM \\
\hline
Yeast\_SPO &  4979 &  14 & 0.8565 & 0.8525 & 0.8687 & 0.8535 \\
Yeast\_SPO &  4979 &  28 & 0.8312 & 0.8263 & 0.8432 & 0.8221 \\
Yeast\_SPO &  4979 &  42 & 0.8131 & 0.8012 & 0.8436 & 0.7987 \\
Yeast\_SPO &  4979 &  56 & 0.8031 & 0.7845 & 0.8380 & 0.7845 \\
Yeast\_SPO &  4979 &  70 & 0.7871 & 0.7688 & 0.8433 & 0.7659 \\
\hline
Yeast\_SCE &  5890 &  15 & 0.8389& 0.8367 & 0.8544 & 0.8569 \\
Yeast\_SCE &  5890 &  30 & 0.8156& 0.8076 & 0.8297 & 0.8120 \\
Yeast\_SCE &  5890 &  45 & 0.7996& 0.7899 & 0.8207 & 0.7832 \\
Yeast\_SCE &  5890 &  60 & 0.7865& 0.7692 & 0.8165 & 0.7638 \\
Yeast\_SCE &  5890 &  75 & 0.7778& 0.7562 & 0.8181 & 0.7469 \\
\hline
Fly\_DME   & 13190 &  23 & 0.7703 & 0.7758 &    OOM & 0.7929 \\
Fly\_DME   & 13190 &  46 & 0.7035 & 0.7107 &    OOM & 0.7047 \\
Fly\_DME   & 13190 &  69 & 0.6693 & 0.6722 &    OOM & 0.6717 \\
Fly\_DME   & 13190 &  92 & 0.6414 & 0.6432 &    OOM & 0.6483 \\
Fly\_DME   & 13190 & 115 & 0.6362 & 0.6241 &    OOM & 0.6299 \\
\hline
Worm\_CEL  & 17519 &  26 & 0.7535 & 0.7612 &    OOM & 0.7820 \\
Worm\_CEL  & 17519 &  52 & 0.6900 & 0.6977 &    OOM & 0.6934 \\
Worm\_CEL  & 17519 &  78 & 0.6521 & 0.6571 &    OOM & 0.6592 \\
Worm\_CEL  & 17519 & 104 & 0.6271 & 0.6246 &    OOM & 0.6320 \\
Worm\_CEL  & 17519 & 130 & 0.6091 & 0.6042 &    OOM & 0.6153 \\
\hline
Human\_HSA & 20967 &  29 &  0.6728 & 0.6710  &  OOM & 0.7100 \\
Human\_HSA & 20967 &  58 & 0.6140  & 0.6181  &  OOM & 0.6293 \\
Human\_HSA & 20967 &  87 & 0.5858  & 0.5874  &  OOM & 0.6007 \\
Human\_HSA & 20967 & 116 & 0.5680  & 0.5678  &  OOM & 0.5810 \\
Human\_HSA & 20967 & 145 & 0.5513  & 0.5518  &  OOM & 0.5681 \\
\hline
\end{tabular}
\caption{MSE values obtained by all four algorithms on biological network data-sets. OOM indicates that the algorithm ran out of memory and did not complete even a single iteration.}
\label{tab:species}
\end{table}

\begin{figure}[h]
    \centering
    \includegraphics[width=\textwidth, trim={9.5cm 1cm 10.5cm 3cm}, clip]{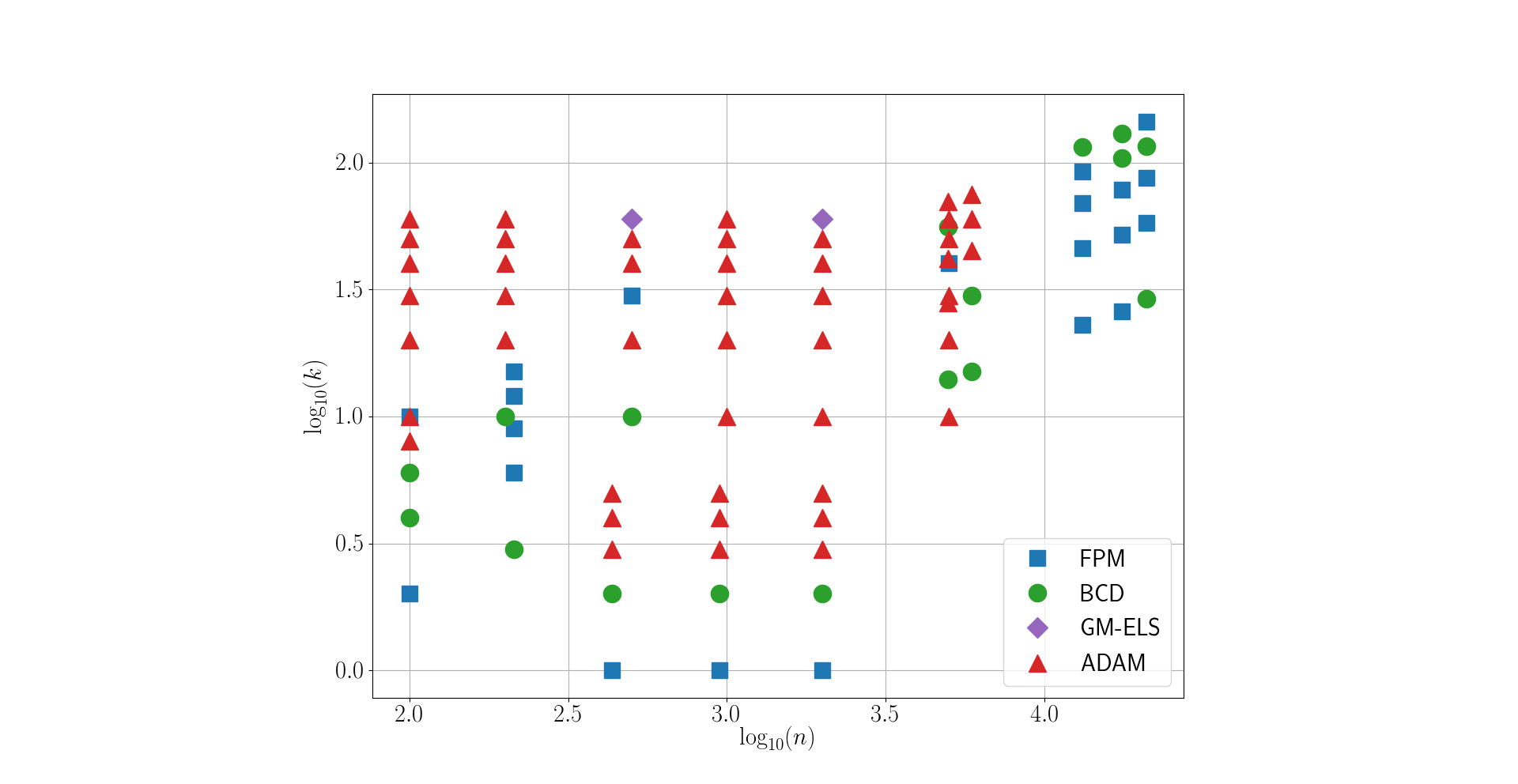}
    \caption{Plot shows which algorithm achieved the lowest MSE of all four algorithms considered with respect to the matrix dimension $n$ and inner dimension $k$. Plot includes  synthetic problems with $K=50$, all  real-world similarity instances   and all biological network instances.
    }
    \label{fig:whowon}
\end{figure}

We summarise the numerical results from Tables \ref{tab:synthetic}--\ref{tab:species} in Figures \ref{fig:MSE_k}--\ref{fig:whowon}.
The former figure shows how $\MSE$ changes (decreases) with increasing inner dimension $k$.
For example, the first 6 plots from the top of the figure depict the results   from Table \ref{tab:synthetic}, where on $x$ axis we put $100k/K$, which has values $20,40,60,100,120$.

We can clearly see that the $\MSE$  decreases with increasing the inner dimension $k$, the only exception is Zonker data-set, where $\MSE$ are very small at the very beginning (below $10^{-2}$) and due to the stopping criteria $\MSE$ does not decrease any further, actually it even slightly increases for the case of FPM, but remains below $10^{-2}$.

To provide better insight which algorithms performs best on given instance (regarding the $\MSE$) we decided to depict the winners in Figure \ref{fig:whowon}. More precisely, for each  synthetic random instance with $K=50$,  each  real-world similarity instance   and each biological network instance we plot  at position $(\log(n),\log(k))$ the  symbol denoting which out of the four methods gives lowest $\MSE$ for given instance within this numerical evaluation.
The more signs appear in this plot the more often corresponding method returns the lowest $\MSE$. 
We omit the synthetic instances with $K<50$  for the sake of more transparent figure.
We can observe that ADAM the most often (in approximately 50\% of the cases) gives the best $\MSE$, followed by BCD and FPM, each winning in approximately 24\% of the instances. The GM-ELS results with best $\MSE$ only in two cases. Additionally, ADAM outperforms the other algorithms for larger values of the inner dimension $k$, while for the smaller values of $k$  the winners are usually FPM and BCD.

\subsection{$\MSE$ vs. time efficiency}
\label{sect:mse_vs_time}

In this subsection we report numerical results where we compare how the $\MSE$ is decreasing with time. We decide to take 3 problem instances, one from each of the three data-sets introduced in Section \ref{sec:data}.
We took only 3 instances to have more transparent figure, but based on our overall observations  we claim that these results are representative for all the three data-sets and all four algorithms and show  the preceding results in a different perspective.

Results clearly indicate that the methods which end with the lowest $\MSE$ on Figure \ref{fig:MSE_k} also spend much more time to reach these results. If we limit the time, which is a usual situation, then FPM method outperforms the others since it demonstrates the best convergence at the beginning and computes the lowest $\MSE$ in a given (short) time-frame.

\begin{figure}
    \centering
    \includegraphics[width=0.85\textwidth]{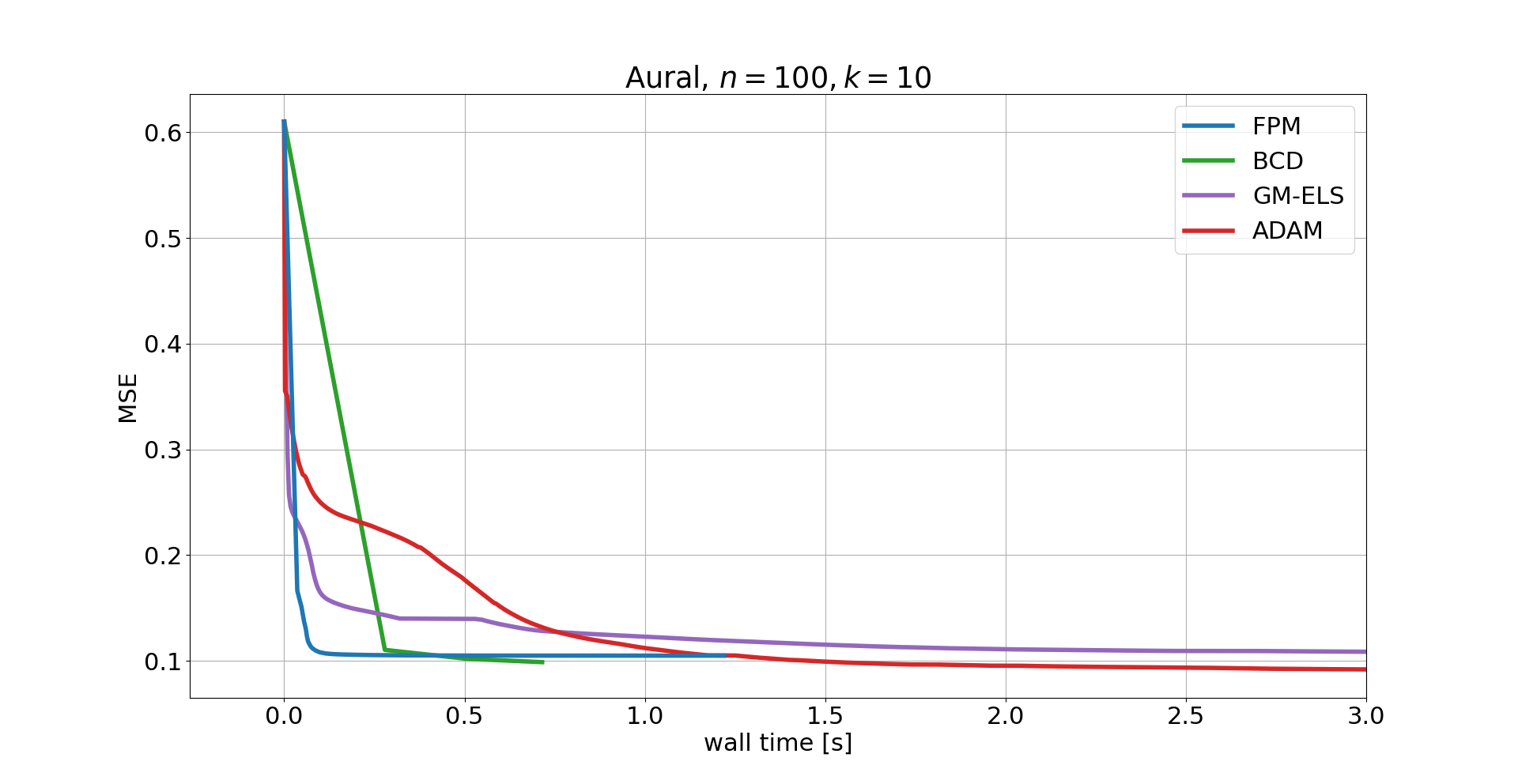}
    \includegraphics[width=0.85\textwidth]{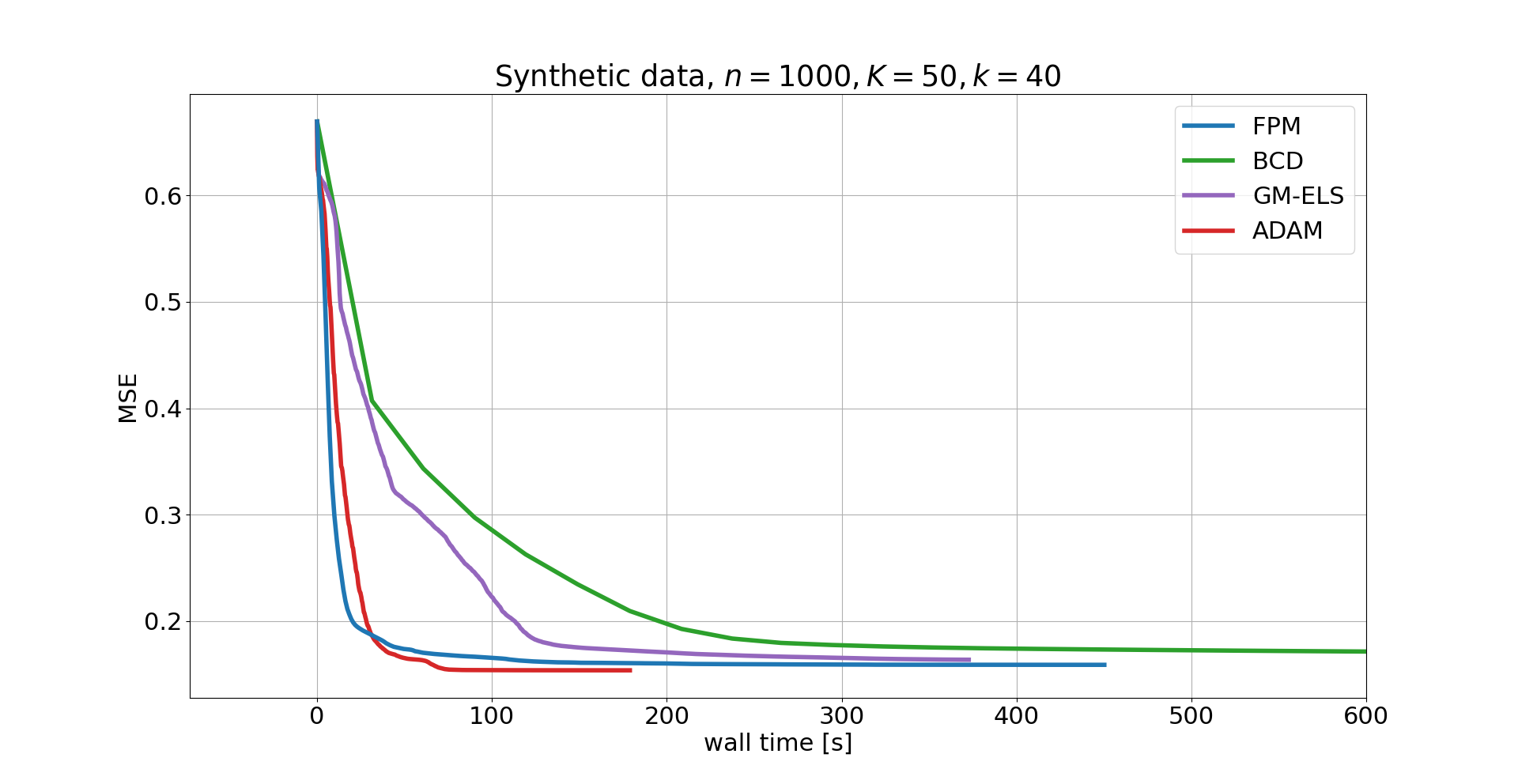}
    \includegraphics[width=0.85\textwidth]{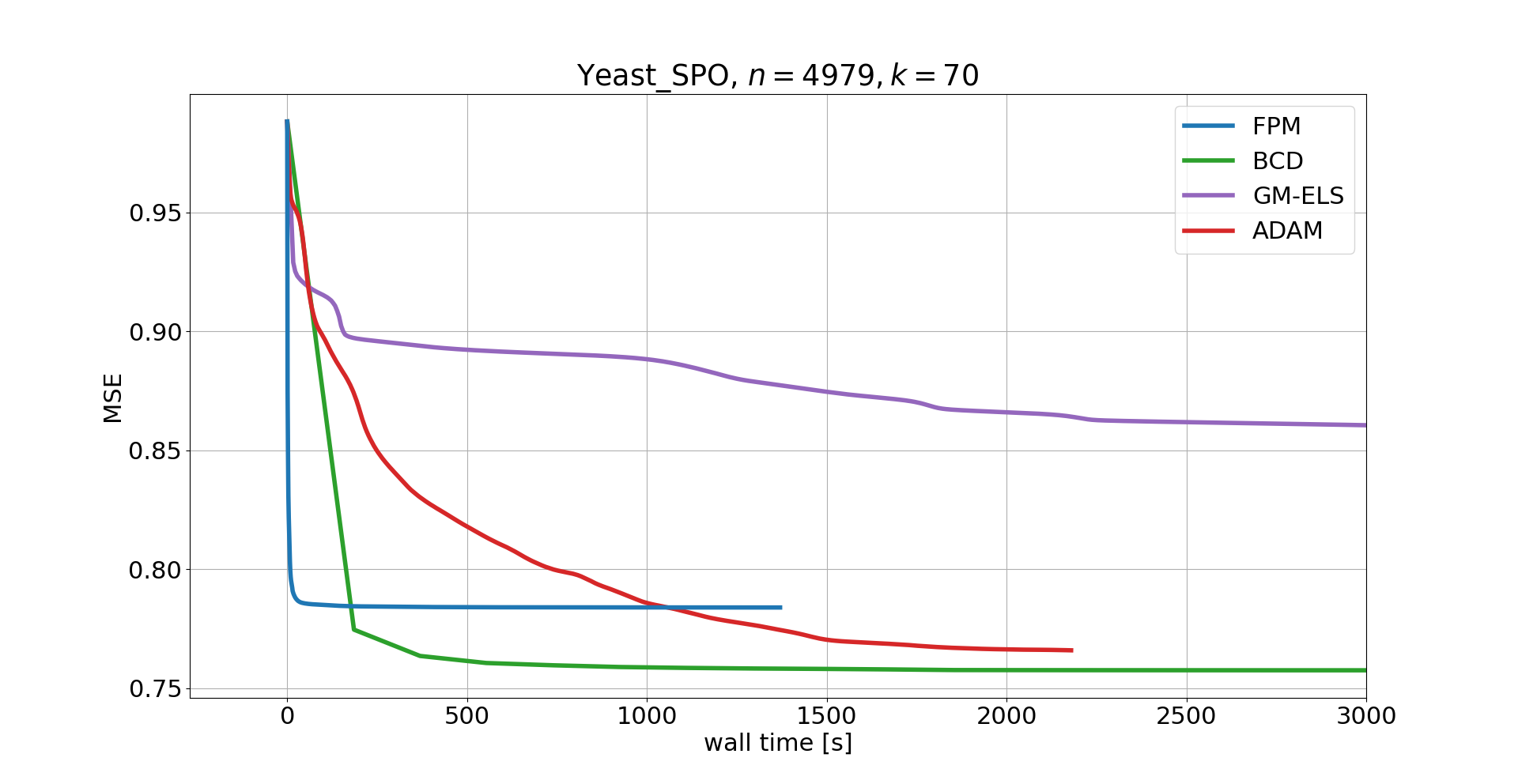}
    \caption{Convergence of all four algorithms considered for three problems with different dimension $n$.}
    \label{fig:progress}
\end{figure}

\section{Discussion and Conclusion}
\label{sec:discussion}

In this paper we developed four algorithms to solve a special variant of non-negative matrix factorization, called symmetric multi-type non-negative matrix tri-factorization problem    \eqref{eqn:SNMTF}, which appears naturally in multi-source data fusion but has not attracted many attention so far.

The first algorithm was an adaptation of the fixed point method which is a classical approach in solving non-negative matrix factorization problems.
The second algorithm was an adaptation of the block coordinate decent method combined with projected gradient method and exact line search, which has already been used in non-negative matrix factorization, but not to solve the \ref{eqn:SNMTF}.

The third and the fourth method were  based on so-called search space transformation. We substituted the factorization variables in such a way that the non-negative constraint vanished at the expense of the cost function.
The gradient method with exact line search (GM-ELS) was based on a substitution of variables by their component-wise squares, while adaptive moment estimation method (ADAM) substituted the variables by their component-wise absolute values. 
In the former case,  the objective function $\SE$ became a multivariate polynomial of order 12 while in the latter case, $\SE$ involved absolute values.


We implemented all the four algorithms into efficient code. The first two algorithms were implemented in Matlab, while the last two in Python with TensorFlow library.
They are available on our GitLab profile \cite{git}.

All the four codes were extensively tested on three data-sets: the first was created by us  and consists of synthetic data for which we know the optimum solution and wanted to reconstruct it back.
We ran all four algorithms for inner dimensions equal to  $k=0.2K, 0.4K,\ldots ,1.2K$, where $K$ was real inner dimension. We  observed that for  $k<1.0K$ the  resulting $\SE$ is decreasing almost linearly with $k$, for all four algorithms, while for $k=K$ and $k=1.2K$ only FPM and ADAM manage to decrease $\MSE$ below $10^{-2}$ in all cases.

The second set consists of real-world data-sets from machine learning and contains 5 real-world similarity or proximity matrices from different fields of data science. 
We used the background description of each data matrix to detect its inner dimension $K$ and then ran all four algorithms with inner dimension $k=0.2K,~0.4K,~\ldots,1.0K$.
We observed that ADAM returned the smallest $\SE$ for all cases while the other algorithms were also performing well, but not on all 5 data matrices.

The third data-set consists of 4 molecular interaction networks for 5 species: 
fission yeast  (Yeast\_SPO), baker's yeast (Yeast\_SCE), fruit fly (Fly\_DME), 
nematode worm (Worm\_CEL) and human (Human\_HSA).
We ran again all the four algorithms with the inner dimensions equal to $k=\round{\sqrt{n}/5},2\round{\sqrt{n}/5},\ldots,5\round{\sqrt{n}/5}$.
We observed that FPM, BCD, and ADAM perform similarly, regarding the $\MSE$, while GM-ELS, on the other hand, is consistently worse than the others or, for larger data-sets, even runs out of memory.

We also analysed  how $\MSE$ was decreasing with time.
We took 3 problem instances, one  from each of the three data-sets.
The results we obtained  clearly indicate that if  we reasonably limit the time, then FPM method performs  best since it demonstrates the best convergence at the beginning and computes the lowest $\MSE$ in a given (short) time-frame. However, if we can afford longer computation times, BCD and ADAM become also very competitive or even better regarding the $\MSE$.

Our findings additionally explain why FPM is so popular in solving non-negative matrix factorization problem. It is indeed very good compromise between convergence quality, time complexity and coding complexity. With very little coding effort and small computing power we can obtain very good feasible solutions. On the other hand, ADAM as a very advanced algorithm, combined with very advanced computing library TensorFlow, demands  much more programming skills and stronger hardware, but often produces better feasible solutions (regarding the $\MSE$), if we provide enough computation time.

We continue our research in solving \ref{eqn:SNMTF} by considering (i) other methods, like 
alternating direction method of multipliers (ADMM), (ii) by  extensions of 
\ref{eqn:SNMTF} which include e.g. orthogonality constraints, and (iii) by considering  theoretical guaranties for the convergence for the methods we are using.

\begin{acknowledgements}
The authors acknowledge the financial support from the Slovenian Research Agency (research core funding No. P2-0098, and projects No. J1-8155,  No. PR-07606, No. N1-0071),  from the  European Research Council (ERC) Consolidator Grant (grant number 770827) 
and from the Spanish State Research Agency AEI 10.13039/501100011033 (grant number PID2019-105500GB-I00). 
\end{acknowledgements}

%
\section*{Conflict of interest}
The authors declare that they have no conflict of interest.

\bibliographystyle{spmpsci}      

\section*{Data availability}
All the data-sets used in this paper and the  Matlab and Python codes for the algorithms presented in this paper  are available on our GitLab portal \cite{git}. 

\bibliography{NMTF_joint_paper.bib}

\appendix

\section*{Appendix}

\section{Calculation of gradient}
\label{sec:grad}

Here we offer a derivation of  \eqref{math:grad1} and \eqref{math:grad2} presented
in Section \ref{ch:gradcalc}. We use notation $[X]_{\mu\nu}$ for component of matrix
$X$ in row $\mu$ and column $\nu$ and $\delta_{ij}$ for the Kronecker delta. Objective
function being differentiated is
\begin{equation*}
    \SE = \sum_{i=1}^N  \| R_i - f(\tilde{G})f(\tilde{S}_i)f(\tilde{G})^\top \|^2 
    = \sum_{i=1}^N \|\D_i\|^2 \\
    = \sum_{i=1}^N \sum_{\mu,\nu} \left[\D_i\right]_{\mu\nu}^2, \\
\end{equation*}
where $\D_i$ is defined in \eqref{eq:Di}.
Let us differentiate $\SE$ with respect to a single component of matrix $\tilde{S}_i$.
\begin{align*}
    \frac{\partial\SE}{\partial [\tilde{S}_i]_{\rho\sigma}} &= \sum_{i=1}^N \sum_{\mu,\nu}\frac{\partial[\D_i]_{\mu\nu}^2}{\partial[\tilde{S}_i]_{\rho\sigma}} 
    = -2\sum_{i=1}^N \sum_{\mu,\nu}[\D_i]_{\mu\nu}\frac{\partial\left[f(\tilde{G})f(\tilde{S}_i)f(\tilde{G})^\top\right]_{\mu\nu}}{\partial[\tilde{S}_i]_{\rho\sigma}} \\
    &= -2\sum_{i=1}^N\sum_{\mu,\nu}[\D_i]_{\mu\nu}\frac{\partial}{\partial[\tilde{S}_i]_{\rho\sigma}} \sum_{p,r}f\left([\tilde{G}]_{\mu p}\right)f\left([\tilde{S}_i]_{pr}\right)f\left([\tilde{G}]_{\nu r}\right) \\
    &= -2\sum_{i=1}^N\sum_{\mu,\nu}[\D_i]_{\mu\nu} \sum_{p,r}f\left([\tilde{G}]_{\mu p}\right)f\left([\tilde{G}]_{\nu r}\right)f'\left([\tilde{S}_i]_{pr}\right)\delta_{\rho p}\delta_{\sigma r} \\
    &= -2\sum_{i=1}^N\sum_{\mu,\nu}[\D_i]_{\mu\nu} f\left([\tilde{G}]_{\mu\rho}\right)f\left([\tilde{G}]_{\nu\sigma}\right)f'\left([\tilde{S}_i]_{\rho\sigma}\right) \\
    &= -2\sum_{i=1}^Nf'\left([\tilde{S}_i]_{\rho\sigma}\right)\left[f(\tilde{G})^T\D_if(\tilde{G})\right]_{\rho\sigma}
\end{align*}
This result can be written more compactly as
\begin{equation*}
    \d S_i~=~\grad_{\tilde{S}_i}\SE = -2\sum_{i=1}^Nf'(\tilde{S}_i)\odot\left(f(\tilde{G})^T\D_if(\tilde{G})\right),
\end{equation*}
which proves \eqref{math:grad1}. 

Using the same steps for $\tilde{G}$, we obtain \eqref{math:grad2}.

\section{Calculation of polynomial used in the exact line search}
\label{sec:poly}

Here we show exactly how the polynomial \eqref{math:poly} used by GM-ELS is calculated.
This polynomial tells how $\SE$ changes with respect to step size $t$ when
making a move in direction of gradient. For sake of brevity we use notation 
$X^2=X\odot X$.
\begin{equation*}
    p(t)=\sum_{i=1}^N\|R_i-(\tilde{G}+t\d \tilde{G})^2(\tilde{S}_i+t\d \tilde{S})^2(\tilde{G}+t\d \tilde{G})^{2\top}\|^2
    =\sum_{j=0}^{12} c_jt^j
\end{equation*}
To calculate coefficients $c_j$ we first express how matrix $\D_i$ changes
with respect to $t$.
\begin{equation*}
    \D_i(t)=R_i-(\tilde{G}+t\d \tilde{G})^2(\tilde{S}_i+t\d \tilde{S}_i)^2(\tilde{G}+t\d \tilde{G})^{2\top}=\sum_{j=0}^6 A_{ij}t^j,
\end{equation*}
where matrices $A_{ij}$ can be calculated using simple expansion steps and are equal to
\begin{align*}
    A_{i0}&=R_i-\tilde{G}^2\tilde{S}_i^2\tilde{G}^{2\top}\\
    A_{i1}&=-2\tilde{G}^2\tilde{S}_i^2(\tilde{G}\odot \d \tilde{G})^\top-2\tilde{G}^2(\tilde{S}_i\odot \d \tilde{S}_i)\tilde{G}^{2\top}-2(\tilde{G}\odot \d \tilde{G})\tilde{S}_i^2\tilde{G}^{2\top}\\
\nonumber
    A_{i2}&=-\tilde{G}^2\tilde{S}_i^2\d \tilde{G}^{2\top}-4\tilde{G}^2(\tilde{S}_i\odot \d \tilde{S}_i)(\tilde{G}\odot \d \tilde{G})^\top-\tilde{G}^2\d \tilde{S}_i^2\tilde{G}^{2\top}-\d \tilde{G}^2\tilde{S}_i^2\tilde{G}^{2\top}\\
    &\quad\,-4(\tilde{G}\odot \d \tilde{G})(\tilde{S}_i\odot \d \tilde{S}_i)\tilde{G}^{2\top}-4(\tilde{G}\odot \d \tilde{G})\tilde{S}_i^2(\tilde{G}\odot \d \tilde{G})^\top\\
\nonumber
    A_{i3}&=-2\tilde{G}^2(\tilde{S}_i\odot \d \tilde{S}_i)\d \tilde{G}^{2\top}-2\tilde{G}^2\d \tilde{S}_i^2(\tilde{G}\odot \d \tilde{G})^\top-2(\tilde{G}\odot \d \tilde{G})\tilde{S}_i^2\d \tilde{G}^{2\top}\\
\nonumber
    &\quad\,-8(\tilde{G}\odot \d \tilde{G})(\tilde{S}_i\odot \d \tilde{S}_i)(\tilde{G}\odot \d \tilde{G})^\top-2(\tilde{G}\odot \d \tilde{G})\d \tilde{S}_i^2\tilde{G}^{2\top}\\
    &\quad\,-2\d \tilde{G}^2\tilde{S}_i^2(\tilde{G}\odot \d \tilde{G})^\top-2\d \tilde{G}^2(\tilde{S}_i\odot \d \tilde{S}_i)\tilde{G}^{2\top}\\
\nonumber
    A_{i4}&=-\tilde{G}^2\d \tilde{S}_i^2\d \tilde{G}^{2\top}-4(\tilde{G}\odot \d \tilde{G})(\tilde{S}_i\odot \d \tilde{S}_i)\d \tilde{G}^{2\top}-4(\tilde{G}\odot \d \tilde{G})\d \tilde{S}_i^2(\tilde{G}\odot \d \tilde{G})^\top\\
    &\quad\,-\d \tilde{G}^2\tilde{S}_i^2\d \tilde{G}^{2\top}-4\d \tilde{G}^2(\tilde{S}_i\odot \d \tilde{S}_i)(\tilde{G}\odot \d \tilde{G})^\top-\d \tilde{G}^2\d \tilde{S}_i^2\tilde{G}^{2\top}\\
    A_{i5}&=-2(\tilde{G}\odot \d \tilde{G})\d \tilde{S}_i^2\d \tilde{G}^{2\top}-2\d \tilde{G}^2(\tilde{S}_i\odot \d \tilde{S}_i)\d \tilde{G}^{2\top}-2\d \tilde{G}^2\d \tilde{S}_i^2(\tilde{G}\odot \d \tilde{G})^\top\\
    A_{i6}&=-\d \tilde{G}^2\d \tilde{S}_i^2\d \tilde{G}^{2\top}.
\end{align*}
Knowing matrices $A_{ij}$, the element-wise square of $\D_i(t)$ is 
\begin{equation*}
\D_i^2(t)=\sum_{j=0}^{12}B_{ij}t^j\\
\end{equation*}
with
\begin{equation}
    \label{math:polysquare}
B_{ij}=\sum_rA_{ir}\odot A_{i,j-r}.
\end{equation}
equation~\eqref{math:polysquare} is an application of equivalence of polynomial multiplication
and discrete convolution. Therefore, the sum is over all $r$ values that lead to legal
indices for both $A_{ir}$ and $A_{i,j-r}$ which are all $r$ between $\max(0, j-6)$ and $\min(j, 6)$.
Remembering
\begin{equation*}
    p(t)=\sum_{i=1}^N\sum_{\mu,\nu}\left[\D_i^2(t)\right]_{\mu\nu}=\sum_{j=0}^{12} c_jt^j
\end{equation*}
we can now express
\begin{equation}
\label{math:coeff}
c_j=\sum_{i=1}^N\sum_{\mu,\nu}\left[B_{ij}\right]_{\mu\nu}.
\end{equation}

Calculation of coefficients $c_j$ requires far higher number of matrix multiplications compared to the number needed for gradient calculation. When implemented in TensorFlow calculation of $c_j$ is approximately $7$ times more expensive than gradient calculation.

\end{document}